\documentclass[aps,pre,preprint,groupedaddres,showkeys,nofootinbib]{revtex4-1}
\usepackage[draft]{changes}
\definechangesauthor[name={Luca Berti}, color=blue]{lb}
\newcommand{\BEM}{BEM}
\newcommand{\norm}[1]{||#1||_2}

\newlength{\bracewidth}
\usepackage{subcaption}
\newcommand{\myunderbrace}[2]{\settowidth{\bracewidth}{$#1$}\hspace*{-1\bracewidth}\smash{\underbrace{\makebox{$#1$}}_{#2}}}

\usepackage[utf8]{inputenc}
\usepackage[english]{babel}
\usepackage{amsmath,amsfonts,amssymb,amstext}
\usepackage{graphicx}
\usepackage{hyperref}
\usepackage{siunitx}
\usepackage{pgfplots}

\usepackage{tikz}
\graphicspath{{figures/}}
\usepackage{parskip}
\usepackage[normalem]{ulem}
\usetikzlibrary{matrix}
\usetikzlibrary{decorations.pathmorphing}
\usetikzlibrary{decorations.markings}
\usetikzlibrary{arrows.meta,bending}
\usepackage{tkz-euclide}
\begin{document}
	\def\radius{0.08\textwidth}
\def\smallradius{0.008\textwidth}

\title{Shapes enhancing the propulsion of multiflagellated helical microswimmers}

\author{Luca Berti$^{1}$}
\email{berti@math.unistra.fr}
\author{Mickaël Binois$^{2}$}
\author{François Alouges$^{3}$}
\author{Matthieu Aussal$^{3}$}
\author{Christophe Prud'Homme$^{1}$}
\author{Laetitia Giraldi$^{2}$}
\email{laetitia.giraldi@inria.fr}
\affiliation{$^1$Cemosis, IRMA UMR 7501, CNRS, Université de Strasbourg, France \\
$^2$Université Côte d’Azur, Inria, CNRS, France \\
$^3$CMAP, Centre de Mathématiques Appliquées Ecole Polytechnique, France}

\date{\today}

\begin{abstract}
	In this paper we are interested in optimizing the shape of  multi-flagellated helical microswimmers.
	Mimicking the propagation of helical waves along the flagella, they self-propel by rotating their tails. The swimmer's dynamics is computed using the Boundary Element Method, implemented in the open source Matlab library \emph{Gypsilab}.
	We exploit a Bayesian optimization algorithm to maximize the swimmer's speeds through their shape optimization. 
	Our results show that the optimal tail shapes are helices with large wavelength, such that the shape periodicity is disregarded. Moreover, the best propulsion speed is achieved for elongated heads when the swimmer has one or two flagella. Surprisingly, a round head is obtained when more flagella are considered. Our results indicate that the position and number of flagella modify the propulsion pattern and play a significant role in the optimal design of the head. It appears that Bayesian optimization is a promising method for performance improvement in microswimming.
\end{abstract}

\keywords{multiflagellated helical microswimmer; Bayesian optimization; shape optimization}

\maketitle

\section{Introduction}
Microswimming is a challenging field due to its applications in biology and engineering 
\cite{LaugaPowers2009,Fischer2018,NelsonKaliakatsos10}. 
Improvement of microswimmer's performance has attracted a lot of attention in the recent literature \cite{Faivre2015}. 

When the swimmer has a microscopic size, the regime of interest is characterized by a Low Reynolds number. This regime imposes hydrodynamical obstructions to microswimmers' stroke patterns and shapes due to the time-reversibility of the fluid flow \cite{Purcell1977}. 

Optimization appears in many aspects of microswimming.
Numerous studies address the path-planning and optimal navigation problems \cite{Liebchen_2019,Daddi-Moussa-Ider-2021}, and recently the method of reinforcement learning has been explored to solve them \cite{Bec2020}.
Other works deal with the optimization of the deformation strategy to enhance the swimmer's efficiency \cite{tam2007optimal,Golestanian2019efficiency,Ishimoto2016} or velocity \cite{Yacine2020}, which consists in finding the best cycle of deformation, namely a stroke, to move. 
Many approaches exist as using the Pontryagin principle \cite{or2016}, or equivalently the Euler-Lagrange equations \cite{Alouges2019, Alouges2009}. 
Propulsion at micro-scale depends on body shapes, and optimizing it becomes another crucial issue.
Parametrical studies allow to investigate the effect of geometrical parameters \cite{Gadelha2013,PhanThien1987}. Using shape optimization theory, \cite{Walker2013} optimizes helical swimmers in Stokes flow. Experimental study \cite{Ye2013} allows to improve the speed of helical microrobot considering multi-flagella.
All these optimization problems are very challenging due to the numerical complexity of the swimmer's dynamical system.

This paper focuses on the shape optimization of multi-flagellated helical microswimmer, where both the head shape and the flagella design are adressed to improve the swimmer's speed.
The swimmer's dynamics is solved using the Boundary Element Method (BEM) which has been extensively used in the microswimming field \cite{Pozrikidis02,PhanThien1987,Shum2019,GaffneyIshimoto2019}. In the rest, the BEM library \emph{Gypsilab} is used to solve the fluid-structure interaction \cite{Gypsi2018}. 
The optimization is then performed using Bayesian optimization \cite{williams2006gaussian}, which is a new method in this context. 
Our results show that the larger the wavelength of the helical tail, the greater the propulsion speed. 
Moreover, the best propulsion speed is achieved for elongated heads when swimmers have one or two flagella. 
Surprisingly, a round head is obtained when more flagella are considered. 
Our results indicate that the position and number of flagella modify the propulsion pattern and play a significant role in the optimal design of the head. 

The paper is organized as follows: in Section \ref{Section:MathModel} the mathematical modeling of the swimming problem is introduced; in Section \ref{Section:NumMeth} the numerical methods are presented, namely the Boundary Element Method and the optimization procedure; in Section \ref{Sect:Results} the results are detailed, focusing first on the monoflagellated case, then on the biflagellated case and finally on the tetra-flagellated case. Section \ref{Sect:Discussion} summarizes and concludes the paper.

\section{Mathematical modeling}
\label{Section:MathModel}
\begin{figure}
	\includegraphics[width=0.6\linewidth]{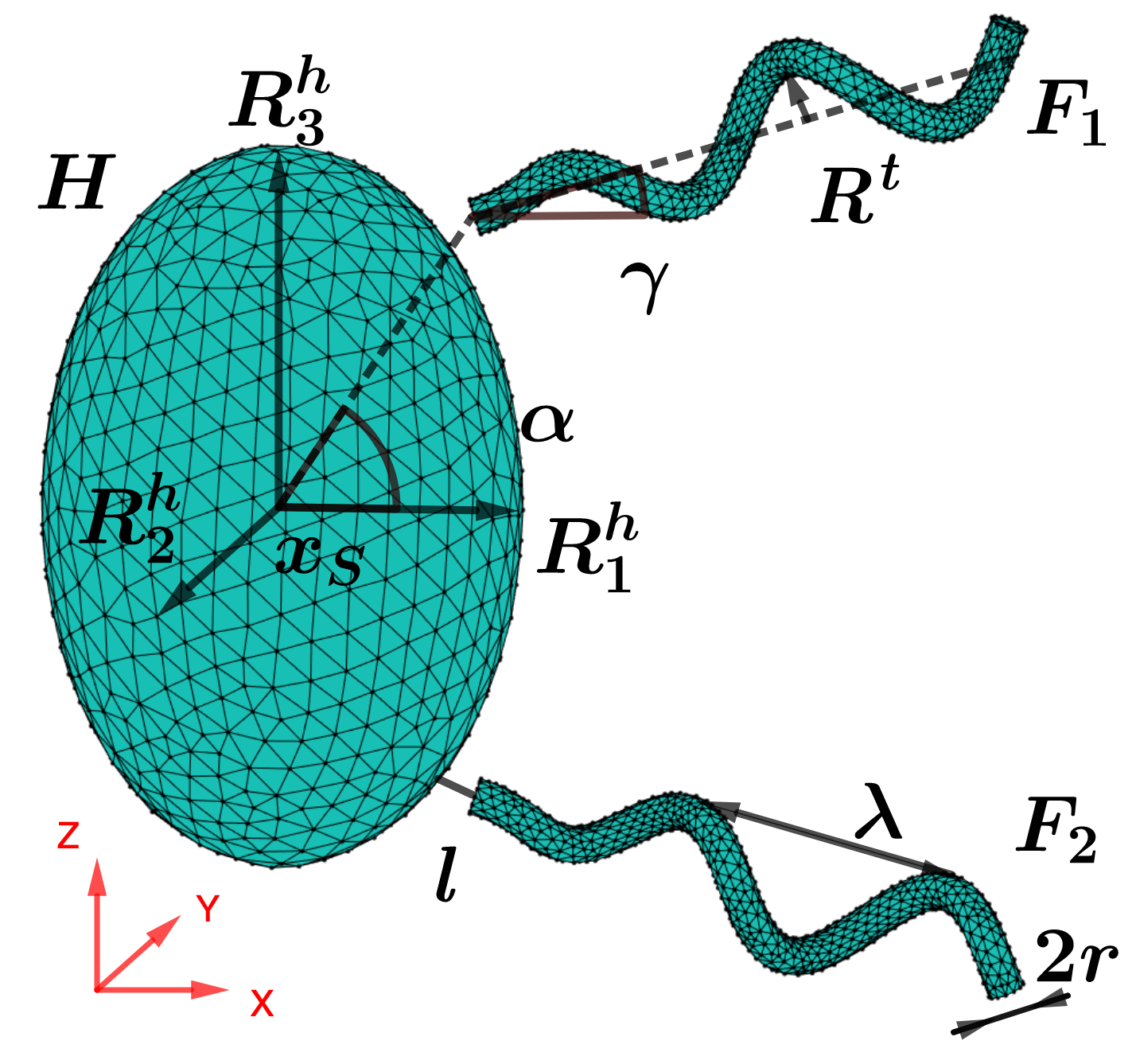}
	\caption{Three-dimensional mesh model of the MO-1 bacterium ($n_T=2$). In this picture the gaps $l$ between the cell body and the tails are visible. The head is an ellipsoid with major axis $R_3^h$ and minor axes $R_1^h=R_2^h$. The tail total length is $L$, its sectional radius is $r$. The helix wavelength is $\lambda$ and its maximal radius is $R^t$. The analytical expression for the cell body and  the tail's centerline are given in \eqref{Eq:HeadShape} and \eqref{Eq:TailShape}, respectively.}
	\label{Fig:Swimmer}
\end{figure}

\paragraph*{\textbf{Swimmer}.}
In this paper we study the shape of three-dimensional self-propelling microswimmers inspired by MO-1 bacteria (see \cite{Lefevre2009,Zhang2014,Shum2019} and figure \ref{Fig:Swimmer} for MO-1). The study will focus on different swimmers having a number of tails $n_T \in \{1,2,4\}$. 
The swimmer $S$ is composed of non-deformable parts: an ellipsoidal head $H$, and $n_T$ helical tails, where $n_T$ varies according to the microswimmer in consideration. The ellipsoidal head  is described by the equation
\begin{equation}
H = \Big\{ (x,y,z): 
	\frac{x^2}{(R_1^h)^2} + \frac{y^2}{(R_2^h)^2} + \frac{z^2}{(R_3^h)^2} = 1 \Big\}
	\label{Eq:HeadShape}
\end{equation}
where $R_1^h$ is the semi-axis in the propulsion direction and $R_2^h,R_3^h$ are the two orthogonal semi-axes.
The tails, denoted by $F_i$, $i=1,...,n_T$, are tubes of radius $r$, having as centerline the curve of total length $L$ described by
\begin{equation}
	\begin{aligned}
	x(s) &= s, \\
	y(s) &= R^t(1-e^{-k_E^2s^2})\cos(2\pi s /\lambda), \\
	z(s) &= R^t(1-e^{-k_E^2s^2})\sin(2\pi s/\lambda),
	\end{aligned}
	\label{Eq:TailShape}
\end{equation}
where $R^t$ is their maximal radius, $\lambda$ is their wavelength and $k_E$ is a shrinkage coefficient \cite{Higdon1979}. 
Tails are separated from the cell body by a small gap $l$, measured along the normal to the ellipsoid, and are symmetrically distributed and rotated with respect to the propulsion direction. The latitude of the tail junctions is denoted by $\alpha$ while their inclination angle with respect to the horizontal is indicated by $\gamma$. The previous notations are presented in Figure \ref{Fig:Swimmer}, in the case of a biflagellated swimmer $(n_T=2)$. In order to swim, the helices rotate around their axes at speed $\omega=\SI{-2\pi}{\radian \second^{-1}} $, mimicking bacteria which propagate helical waves along their tail. This modelisation was already employed in \cite{Shum2019} for the biflagellated swimmer, and in \cite{PhanThien1987} for the monoflagellated one.
\\
\paragraph*{\textbf{Swimmer's dynamics.}} The fluid is modeled via Stokes equations, due to the small value of the Reynolds number for microswimmers. Fluid velocity and pressure, denoted by $(u,p)$, satisfy the following Dirichlet boundary value problem when the swimmer is composed of one head $H$ and several flagella
\begin{equation*}
\left\{
\begin{aligned}
\nabla p -\mu \Delta u = 0 \quad &\text{on $\mathbb{R}^3\setminus S$}\\ 
\nabla \cdot u = 0	\quad &\text{on $\mathbb{R}^3\setminus S$}\\ 
u = U + \Omega \wedge (x-x_S) \quad &\text{on $\partial H$}\\
u = U + \Omega \wedge (x-x_S) + \omega \vec{e}_1^{F_i}\wedge(x-x^{F_i})\quad &\text{on $\partial F_i$}, \\&\text{ $i=1,..,n_T$}
\end{aligned}
\right.
\label{Eq:Stokes}
\end{equation*}
where $S = H \cup F_1 \cup \dots \cup F_{n_T}$, $x_S$ is the center of mass of the head, $\vec{e}_1^{F_i}$ is the axis direction of the $i-$th tail, for $i\in \{1,\dots, n_T\}$, $x^{F_i}$ is the $i-$th tail's junction $i\in \{1,\dots, n_T\}$, $U\in \mathbb{R}^3$ and $\Omega \in \mathbb{R}^3$ are the linear and angular velocity of the swimmer. Notice that the Dirichlet boundary conditions are composed of two distinct parts: the term $ \omega \vec{e}_1^{F_i}\wedge(x-x^{F_i})$ depends on the rotation rate of the tail, that is a known datum, while term $U + \Omega \wedge (x-x_S)$ contains the linear and angular velocities that result from the interaction between the swimmer and the fluid, which are unknown.

Let $x = (x_1,x_2,x_3)$ and $y=(y_1,y_2,y_3)$ be points in $ \mathbb{R}^3$.
Three-dimensional Stokes equations, being linear in velocity and pressure, admit a tensorial Green kernel denoted by $G_{ij}(x,y)$, for $i,j=1,2,3$.  For free-space, such Green kernel reads as
\begin{equation}
G_{ij}(x,y)=\frac{1}{8\pi\mu}\Big(\frac{\delta_{ij}}{\norm{x-y}}+\frac{(x_i-y_i)(x_j-y_j)}{\norm{x-y}^3}\Big),
\label{Eq:Stokeslet}
\end{equation}
where $\delta_{ij}$ is the Kronecker delta \cite{Pozrikidis1992,Happel1983}. 

The convolution of the Green kernel with fluid surface tensions $f:\partial S \rightarrow \mathbb{R}^3$ participates to
an integral representation of the flow field $u$ for all $x \in \partial S$, given by
\begin{equation}
u_i(x) = -\int_{\partial S} G_{ij}(x,y)f_j(y) \, \mathrm{d}S(y)\\
+ \int_{\partial S}^{PV} \frac{1}{8\pi}u_j(y)T_{jik}(x,y)n_k(y) \, \mathrm{d}S(y).
\label{Eq:Representation_Stokes}
\end{equation}
where $T_{jik}(x,y)=-6(x_i-y_i)(x_j-y_j)(x_k-y_k)/\norm{x-y}^5$, $n_k(y)$ is the $k-$th component of the outward normal to $\partial S$ and the Einstein summation convention is employed  \cite{Pozrikidis1992}. 

Using the integral representation formula \eqref{Eq:Representation_Stokes}, the linear and angular velocities $(U,\Omega)$ and the surface tensions $f$ can be determined via
	\begin{subequations}
		\begin{align}
		\begin{split}
		 U &- (x-x_S) \wedge \Omega + \\ &\int_{\partial S} G(x,y)f(y) \, \mathrm{d}y =  (x-x^F) \wedge \omega \vec{e}_1^F \label{Eq:Swimming1},
		\end{split}\\
		&\int_{\partial S} f(y) \, \mathrm{d}y = 0 \label{Eq:Swimming2}, \\
		&\int_{\partial S} (y-x_S) \wedge f(y) \, \mathrm{d}y = 0 \label{Eq:Swimming3}.
		\end{align}
		\label{Eq:Swimming}
	\end{subequations}
Equation  \eqref{Eq:Swimming1} derives from  \eqref{Eq:Representation_Stokes}: here we exploited the fact that on the boundary of a rigid body the velocity writes as $u(x)=U + \Omega \wedge (x-x_S)$ and that reduces the second integral in \eqref{Eq:Representation_Stokes} to $0$ when $x\in \partial S$ \cite{Pozrikidis1992}. Equations \eqref{Eq:Swimming2}-\eqref{Eq:Swimming3} respectively indicate that net forces and torques over the swimmer are zero. These two equations are usually named ``self-propulsion constraints" and they model self-propelled motions, i.e. those deriving from internal forces and body deformations. Using $(U,\Omega)$ obtained from \eqref{Eq:Swimming}, the resulting trajectory of the swimmer could be computed.
\\
\paragraph*{\textbf{Optimization problem.}} By using the previous hydrodynamical model, we look for the shapes which optimize the swimmer's velocity in the propulsion direction.  In this case, the shape of the swimmer depends on a finite number of parameters $p \in \mathcal{P}$, where $\mathcal{P} \subset \mathbb{R}^d$ is a compact set and $d\in \mathbb{N}^*$ is the number of variable parameters. The cost function is the first component of the average velocity $\bar{U} = \frac{1}{T}\int_{0}^{T}U(t)\, \mathrm{d}t$, where the swimmer's linear velocity is averaged over the tail's rotation period $T$. The volumes of the head $|H|$ and tail $\sum_{i=1}^{n_T} |F_i|$ are fixed and equal to $\mathit{vol}_H$ and $\mathit{vol}_F$.  Thus, the optimization problem of interest reads as
\begin{equation}
	\max_{
		\substack{p \in\mathcal{P} ,\\
			|H| = \mathit{vol}_H, \\
		\sum_{i=1}^{n_T} |F_i|= \mathit{vol}_F}
		}
	-\bar{U}_x.
\end{equation}

\paragraph*{Remark.} In the case of the monoflagellated swimmer, due to the lack of symmetry in its shape, we consider an additional constraint on $\bar{U}_y,\bar{U}_z$ defined by 
\begin{equation}
	|\bar{U}_y|,|\bar{U}_z| \le \varepsilon, \quad \varepsilon>0.
\end{equation}
Different values of $\varepsilon$ are considered, and their effect on the swimmer's optimal shape are discussed.
\section{Numerical methods}
\label{Section:NumMeth}
\label{Subsec:Optimization}
\paragraph*{\textbf{Boundary element method.}}
\label{Subsect:BEM}
The Boundary Element Method (\BEM) is a well-adapted framework for helical microswimmers. It is a popular mesh-based numerical method which allows the simulation of Stokes flow via the integral formulation \eqref{Eq:Representation_Stokes}, and it has been  used in parametric studies of monoflagellated swimmers \cite{PhanThien1987}. Boundary Element Method solves \eqref{Eq:Representation_Stokes} by evaluating numerically the integrals and by regularizing the singular Green kernel when necessary. A possible regularization method consists in semi-analytical integration \cite{Gypsi2018}.  We detail below the numerical formulation of problem  \eqref{Eq:Swimming}.

The components of surface tensions are expanded as $f_j(y) = \sum_{l=1}^N f_j^l\phi^l(y)$, where $\{\phi^l\}_{l=1}^N$ span the scalar continuous $P^1$ finite element space over $\partial S$, and each component of equation \eqref{Eq:Swimming1} is projected on this space, giving
	\begin{subequations}
	\begin{align}
	\begin{split}
	&\int_{\partial S} U_i \phi^k(x) \, \mathrm{d}x- \int_{\partial S} ((x-x_S) \wedge \Omega)_i \phi^k(x) \, \mathrm{d}x+ \\ &\int_{\partial S}\Bigg(\int_{\partial S} G_{ij}(x,y)\sum_{l=1}^N f_j^l\phi^l(y) \, \mathrm{d}y \Bigg) \phi^k(x) \, \mathrm{d}x =
	\\ &\int_{\partial S} (x-x^F) \wedge \omega \vec{e}_1^F \phi^k(x) \, \mathrm{d}x \quad \text{for $k,l=1,\dots,N$} \label{Eq:Swimming-disc1}
	\end{split}\\
	&\int_{\partial S} \sum_{l=1}^N f_j^l\phi^l(y) \, \mathrm{d}y = 0 \quad \text{for $l=1,\dots,N$} \label{Eq:Swimming-disc2} \\
	&\int_{\partial S} (y-x_S) \wedge \sum_{l=1}^N f_j^l\phi^l(y) \, \mathrm{d}y = 0 \quad \text{for $l=1,\dots,N$} \label{Eq:Swimming-disc3}
	\end{align}
	\label{Eq:Swimming-disc}
\end{subequations} 
Since $\partial S = \partial H \cup \partial F_1 \cup \dots \cup \partial F_{n_T}$, integrals can be split and the
 resulting system matrix will have a block structure reading as
\begin{equation*}
\begin{tabular}{r}
\scalebox{0.75}{$3\times N$} $\left\{\lefteqn{\phantom{\begin{matrix} G \end{matrix}}}\right.$\\
\scalebox{0.75}{$3$} $\left\{\lefteqn{\phantom{\begin{matrix} J \end{matrix}}} \right.$\\
\scalebox{0.75}{$3$} $\left\{\lefteqn{\phantom{\begin{matrix} 0 \end{matrix}}} \right.$
\end{tabular}
	\begin{bmatrix}
	\, \, G & J^T & K^T \\
	\, \, J & 0 & 0 \\
	\, \, \, \, \, \, \myunderbrace{K}{3\times N} & \myunderbrace{0}{3} & \myunderbrace{0}{3} 
	\end{bmatrix}
	\begin{bmatrix}
	f \\
	U \\
	\Omega
	\end{bmatrix}
	= 
	\begin{bmatrix}
	I(\omega)\\
	0 \\
	0
	\end{bmatrix},
\end{equation*}
\\
where the matrices $G$, $J$, $K$ and $I(\omega)$ are defined in appendix \ref{Appendix}.
The implementation is done through the Matlab \BEM \, library \emph{Gypsilab}\footnote{https://github.com/matthieuaussal/gypsilab}.
\\
\paragraph*{\textbf{Bayesian Optimization.}}
Optimization of swimmers is carried out by means of Bayesian optimization
(BO), available in the Matlab \texttt{bayesopt} routine. BO provides global
optimization capabilities for problems where evaluations are expensive. Such
problems appear in many fields, ranging from engineering design, physics,
operations research, to hyperparameter optimization in machine learning. The
interested reader is referred to
\cite{shahriari2015taking,gramacy2020surrogates} for a general review of the
methodology. In a nutshell, starting from a small initial design of
experiments, BO works by first considering a surrogate of the initial
black-box function $f$, such as a Gaussian process (see, e.g.,
\cite{williams2006gaussian}). In our case, $f$ is defined as the function which associates for a set of parameters the swimmer's average velocity, i.e.,  $f:p\in\mathcal{P}\mapsto \bar{U}$. This probabilistic model not only provides a
prediction of $f$ over the entire input space but also comes with uncertainty
information on this prediction in the form of a predictive variance. This is
particularly useful to balance between exploitation of promising input regions
where the predictive mean is good and exploration of unknown ones where the
variance is high, through the use of acquisition function. Among other
alternatives, expected improvement (EI) \cite{Mockus1978} is used to select
designs sequentially as proposed by the so-called efficient global
optimization \cite{jones1998efficient}, a canonical BO method. 

\begin{figure}
	\includegraphics[width=0.45\textwidth]{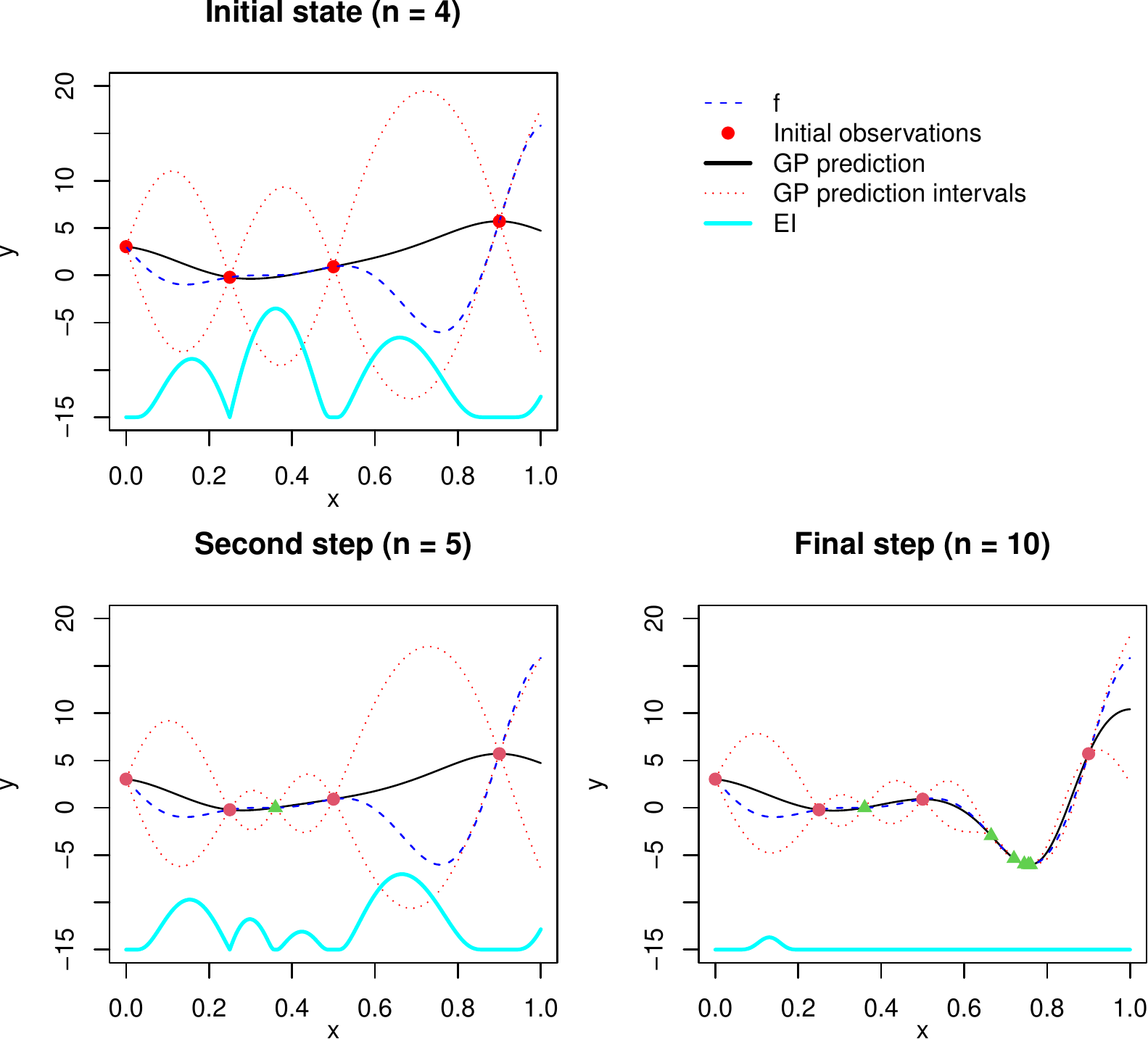}
	\caption{Several steps of Bayesian optimization on a toy function.}
	\label{Fig:EI}
\end{figure}

An illustration of the procedure is proposed in Figure \ref{Fig:EI}. At the
first iteration, the design with maximum EI is selected around $x \approx
0.35$, mislead by a crude initial model. Then the method is able to identify{}
the location where the optimum is located in a very few iterations. The
prediction intervals become smaller as new data comes, and the EI value, of
the same unit as the response, also decreases. This latter can be used to
monitor and, eventually, stop the process.

Dealing with expensive constraints involves modeling them as well. Then they
can be used to weight EI by the corresponding probability of feasibility
\cite{Schonlau1998}. Noise in the responses can be included as well, see for
instance \cite{Letham2018} for application examples.
\begin{center}
	\begin{table}
	\begin{tabular}{c  c  c}
		\hline 
		Symbol & Value (dimensionless) & Value (dimensional) \\
		\hline
		$R_1^h$ & 0.874 & \SI{0.65}{\micro \meter}\\
		$R_2^h$ & 0.874 &  \SI{0.65}{\micro \meter}\\
		$R_3^h$ & 1.5$R_1^h$ &	 \SI{0.975}{\micro \meter}\\
		$L$ & 3.0 &  \SI{2.2}{\micro \meter} \\
		$r$ & 0.067 &  \SI{50}{\micro \meter} \\
		$R^t$ & 0.2 &  \SI{0.15}{\micro \meter} \\
		$\lambda$ & 1.0 &  \SI{0.74}{\micro \meter} \\
		$k_E$ & 0.333$\cdot 2\pi/\lambda = 2.09 $ &  \SI{2.8}{\micro \meter^{-1}} \\
		$l$ & 2$r$ = 0.134 &  \SI{100}{\nano \meter} \\
		\hline
	\end{tabular}
\caption{Parameters describing the body and tail shape of the bacterium. The lengthscale for the adimensionalisation is $\lambda = 0.74$ $\mu m$. This table is taken from \cite{Shum2019}.}
\label{Table:SwimmerParameters}
\end{table}
\end{center}

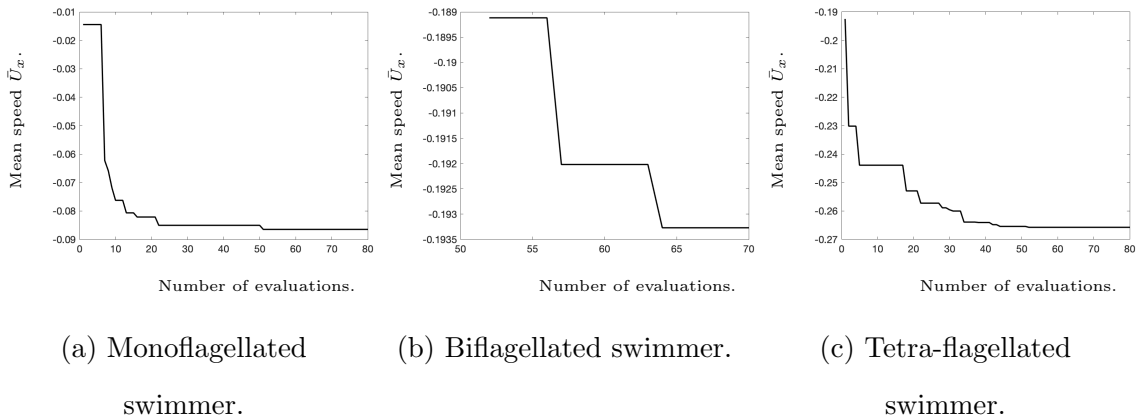
\begin{figure*}
	\begin{subfigure}[t]{.3\linewidth}
		\centering\centering\begin{tikzpicture}
		\node[anchor=south west,inner sep=0] (image) at (0,0){\includegraphics[width=\linewidth]{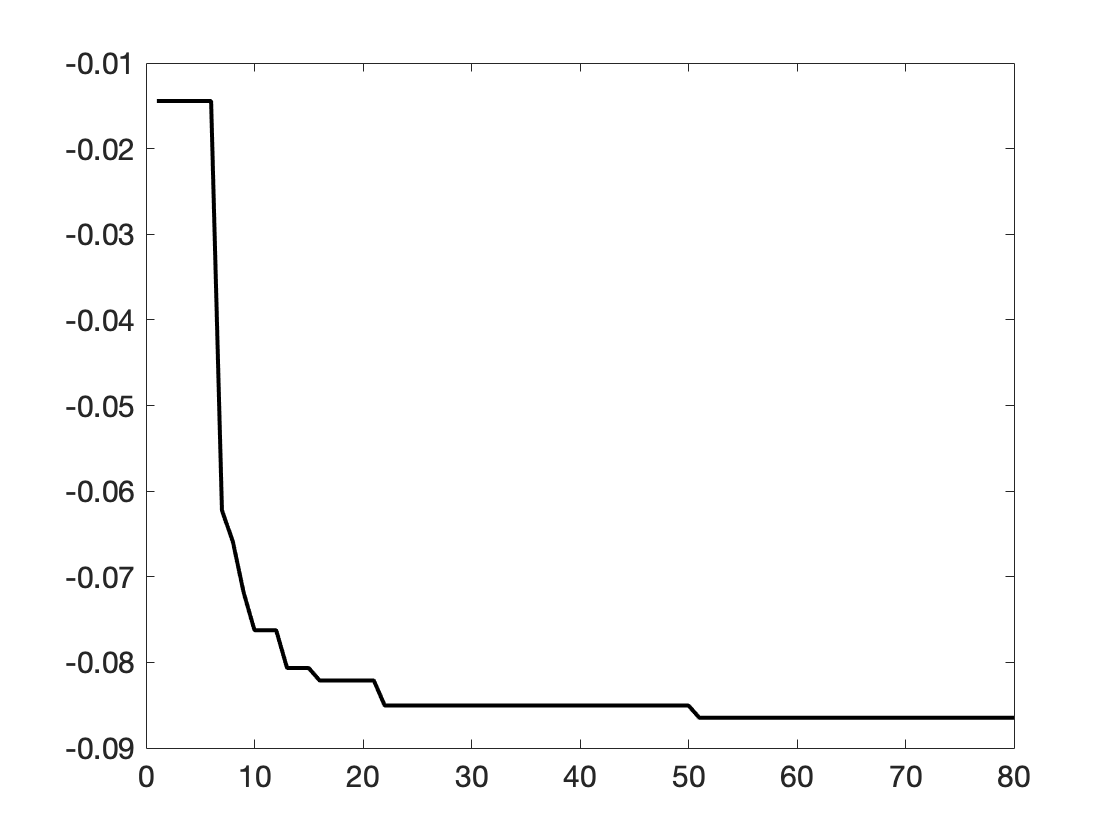}};
		\node[rotate=90] at (-0.2,2) { \tiny Mean speed $\bar{U}_x$.};
		\node at (3,-0.2)  { \tiny Number of evaluations.};
		\end{tikzpicture}
		\caption{Monoflagellated swimmer.}
		\label{Fig: Obj1tail}
	\end{subfigure}
	\begin{subfigure}[t]{.3\linewidth}
		\centering\begin{tikzpicture}
		\node[anchor=south west,inner sep=0] (image) at (0,0){\includegraphics[width=\linewidth]{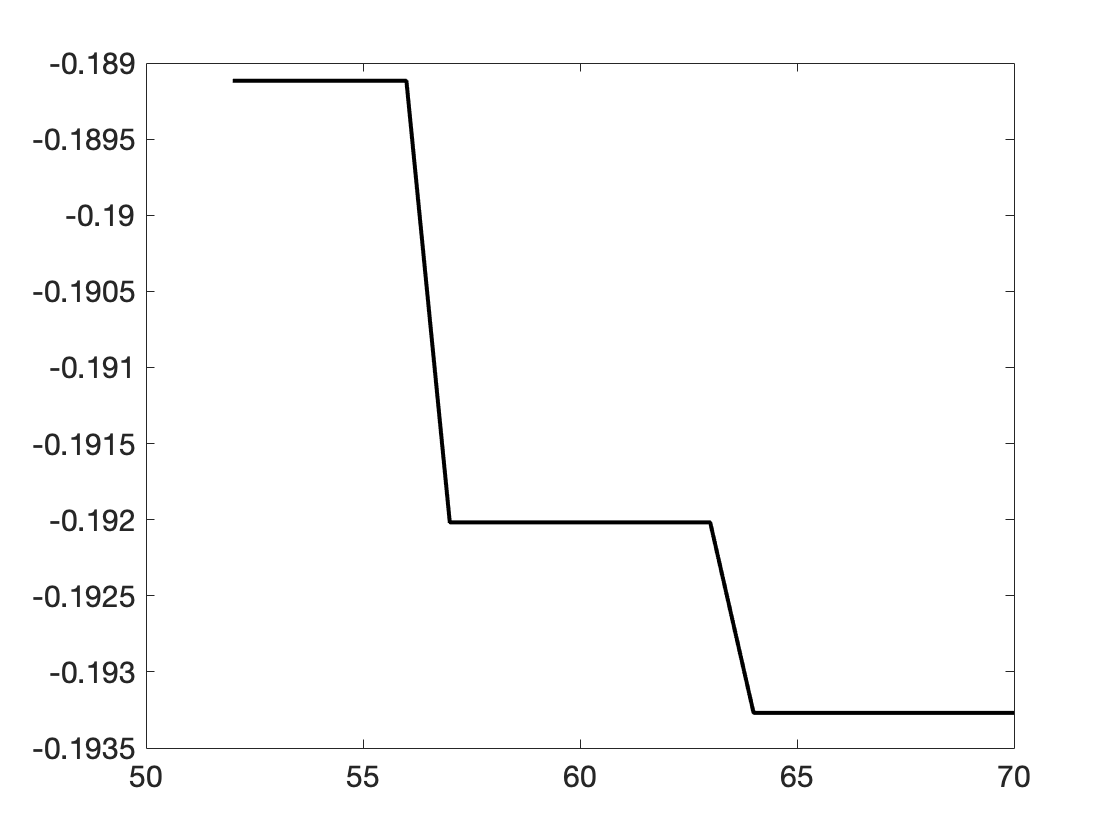}};
		\node at (3,-0.2)  { \tiny Number of evaluations.};
		\node[rotate=90] at (-0.2,2) { \tiny Mean speed $\bar{U}_x$.};
		\end{tikzpicture}
		\caption{Biflagellated swimmer.}
		\label{Fig: Obj2tail}
	\end{subfigure}
	\begin{subfigure}[t]{.3\linewidth}
		\centering\centering\begin{tikzpicture}
		\node[anchor=south west,inner sep=0] (image) at (0,0){\includegraphics[width=\linewidth]{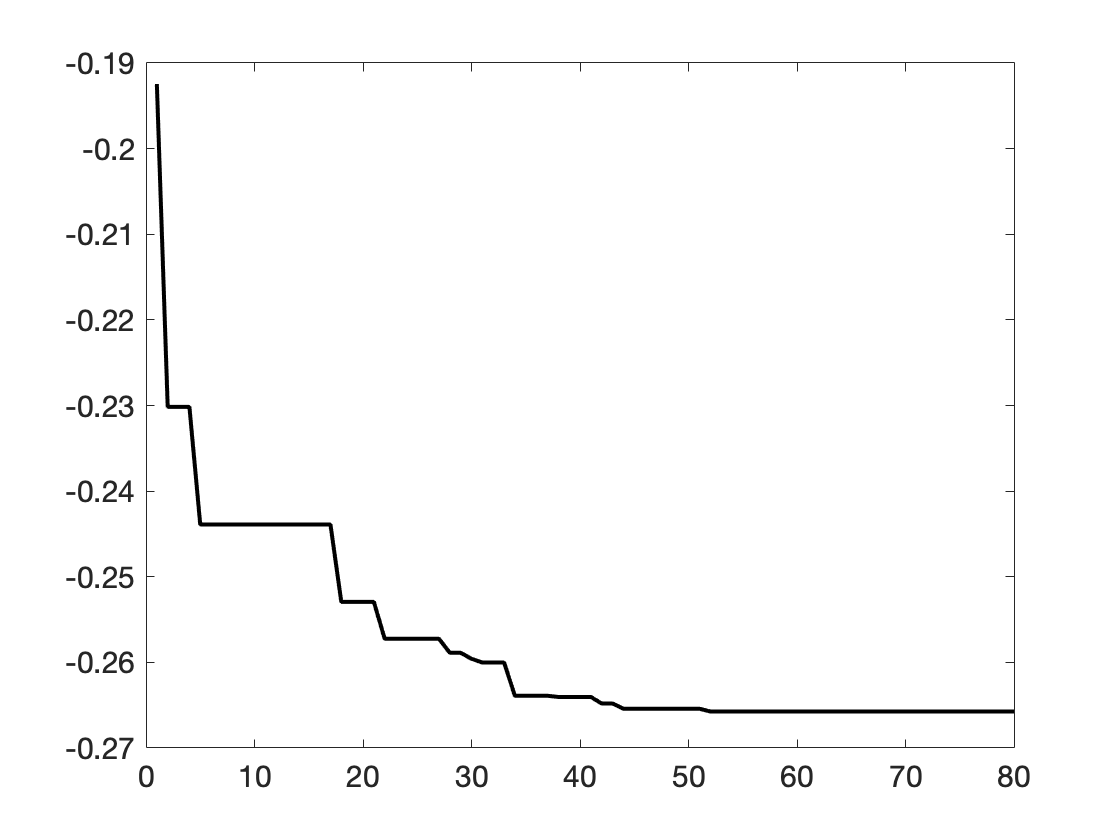}};
		\node at (3,-0.2)  { \tiny Number of evaluations.};
		\node[rotate=90] at (-0.2,2) { \tiny Mean speed $\bar{U}_x$.};
		\end{tikzpicture}
		\caption{Tetra-flagellated swimmer.}
		\label{Fig: Obj4tail}
	\end{subfigure}
	\caption{Behaviour of the mean speed $\bar{U}_x$ versus the number of evaluations realized by the optimization algorithm.}
	\label{Fig: Objtail}
\end{figure*}

\section{Numerical Results}
\label{Sect:Results}
In this section we address the optimization of several swimmers. The values of the geometrical parameters are collected in Table \ref{Table:SwimmerParameters}. The dimensionless quantities are considered, and the variables which are not optimized keep these same values. We first consider a monoflagellated microswimmer and optimize at the same time the tail wavelength, the tail radius and the cell body. 

Secondly, a biflagellated swimmer is considered and, to the previously mentioned parameters, we add two angles defining the placement of the tails. 
The placement of the two tails is symmetric with respect to the propulsion direction (i.e. $x$-axis).

Finally, we carry out the optimization process for a tetra-flagellated swimmer by optimizing  the tail wavelength, the tail radius and the cell body. Also in this setting, the position of the four tails satisfies a symmetry requirement with respect to the propulsion direction.
\subsection{Monoflagellated swimmer}
\label{Subsec: Monoflagellated}
\begin{figure}[h]
	\begin{tikzpicture}
	\node[anchor=south west,inner sep=0] (image) at (0,0) {
		\includegraphics[width=0.33\linewidth]{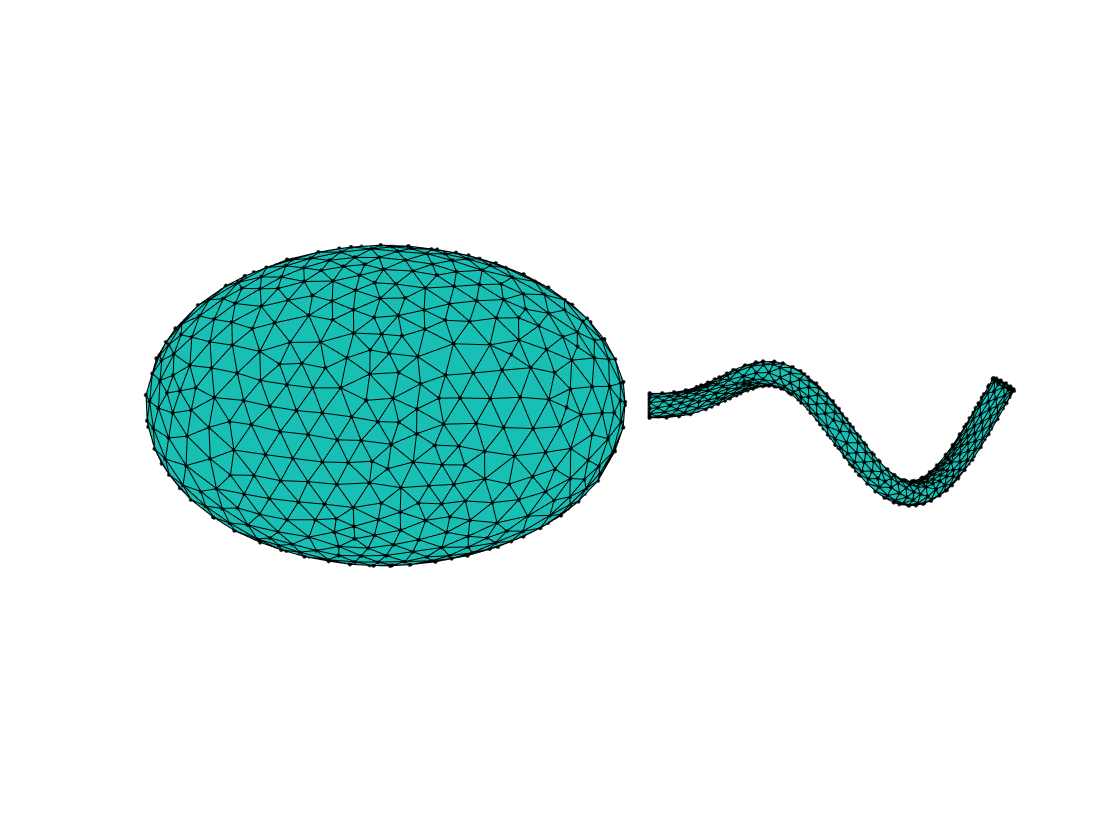}
		\includegraphics[width=0.33\linewidth]{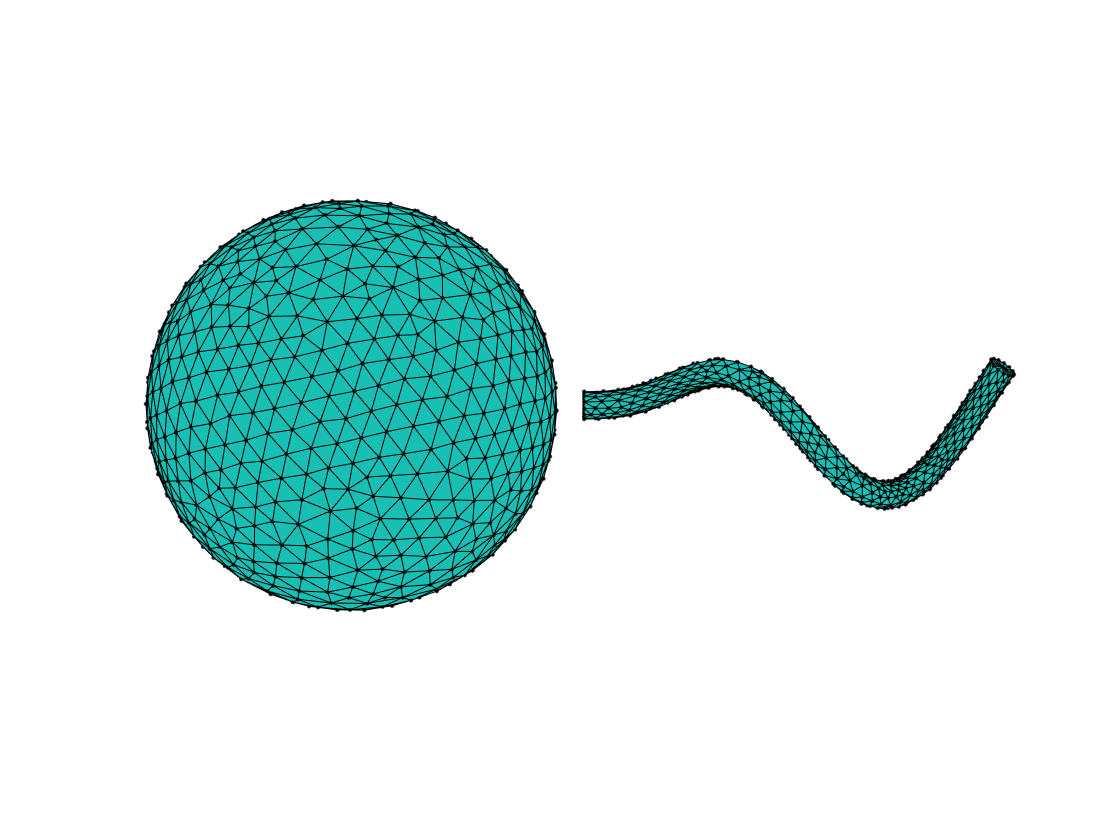}
		\includegraphics[width=0.33\linewidth]{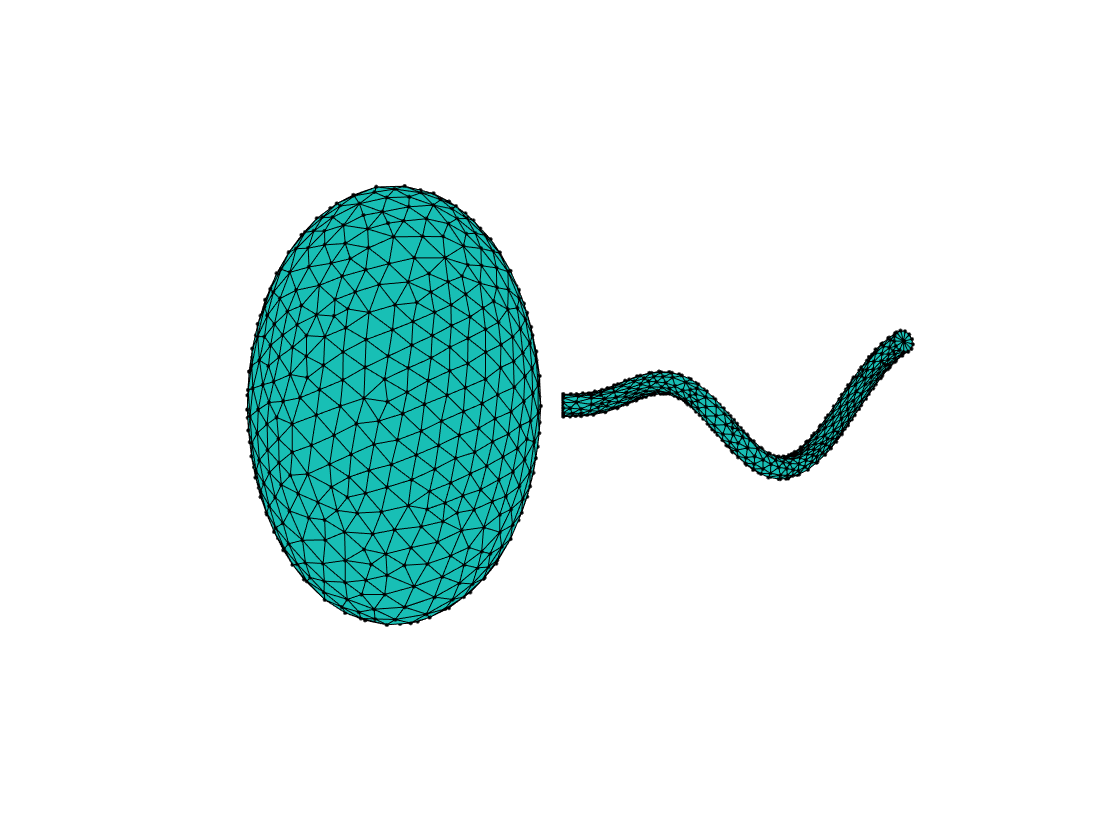}};
	\begin{scope}
	\node at (0.12\textwidth,-0.3) {(A). \textit{Prolate} head.};
	\node at (0.52\textwidth,-0.3) {(B). \textit{Spherical} head.};
	\node at (0.82\textwidth,-0.3) {(C). \textit{Oblate} head.};
	\end{scope}
	\end{tikzpicture}
	\caption{Terminology for the head shapes.}
	\label{Fig:3MONOTAIL}
\end{figure}
The presence of only one helical flagellum can produce a non-negligible transversal motion.
In order to ensure an horizontal self-propulsion, we introduce an additional constraint on the absolute value of the transverse velocities. This allows to produce motions that are prevalently straight. In figure \ref{Fig:3MONOTAIL}, we illustrate three adjectives describing the head shapes as we will use them in this paper: \textit{prolate} ($R_1^h > R_3^h$), \textit{spherical} and \textit{oblate} ($R_1^h < R_3^h$). 

\begin{figure}[h]
	\begin{subfigure}{0.26\linewidth}
\begin{minipage}{\linewidth}
	\includegraphics[width=\linewidth]{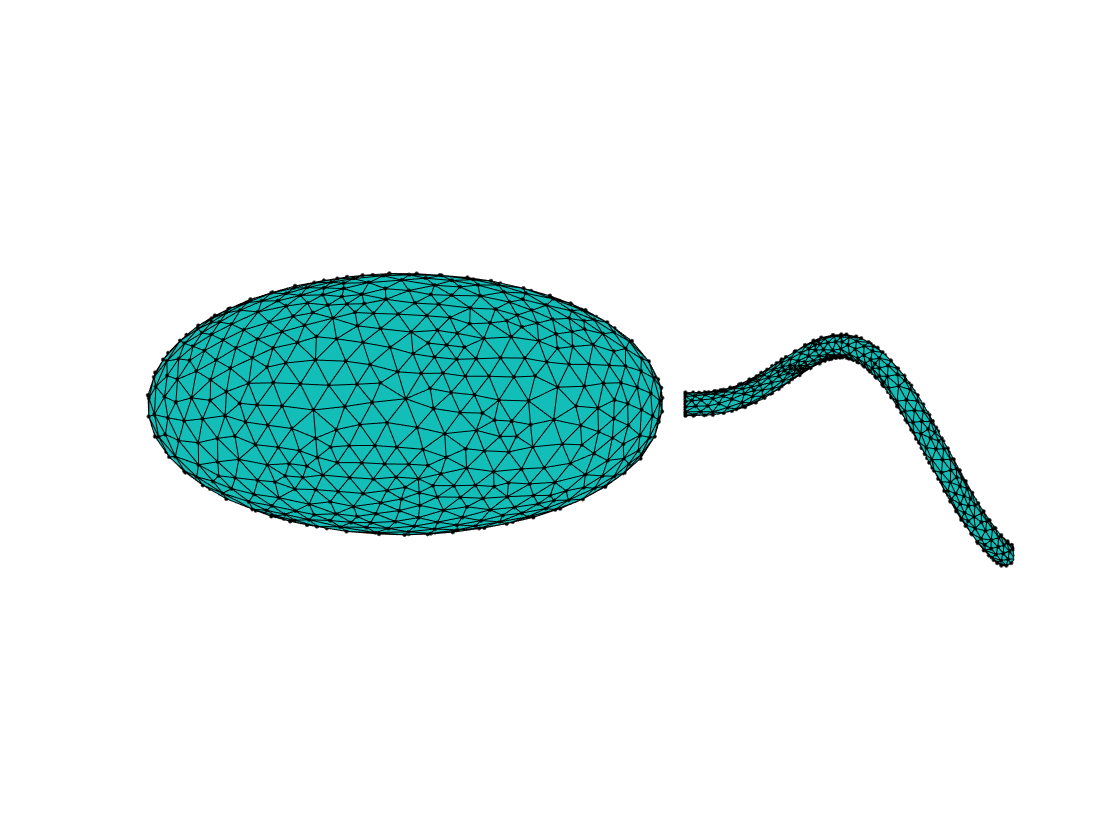}\\
	\includegraphics[width=\linewidth]{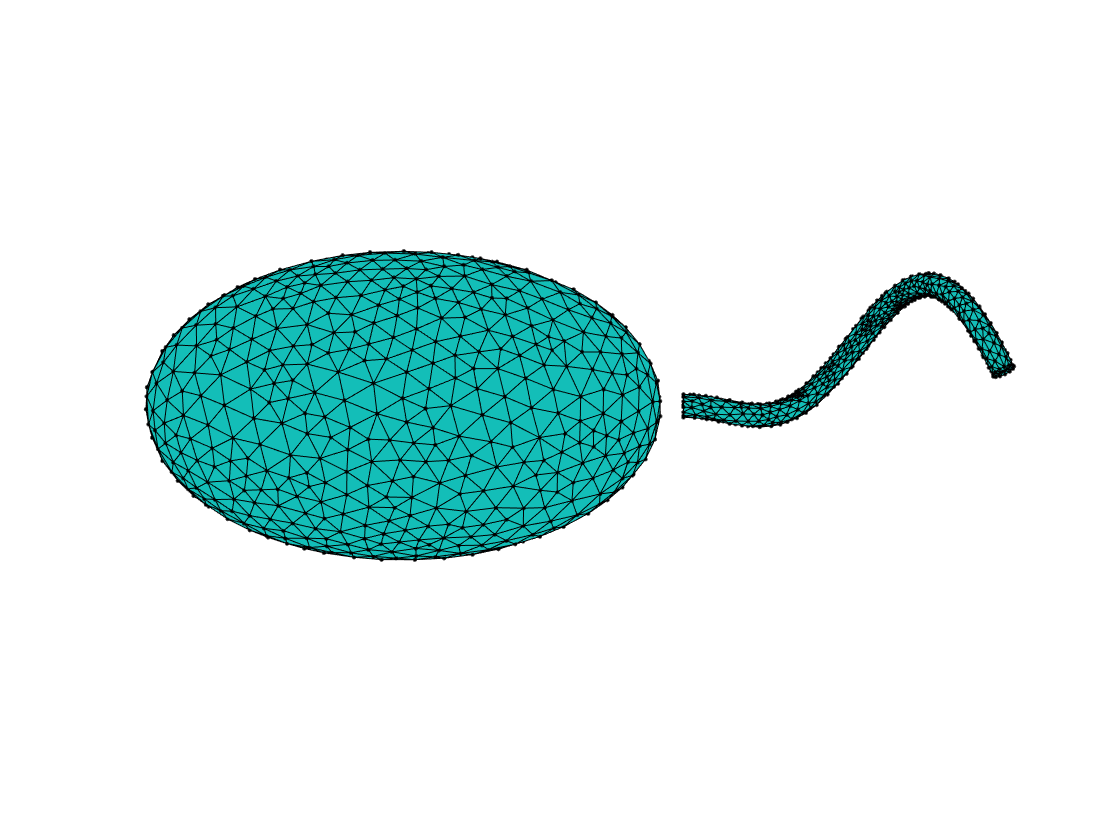}
\end{minipage}
\caption{Top and side view.}
\end{subfigure}
\begin{subfigure}{0.7\linewidth}
\begin{minipage}{\linewidth}
		\pgfplotsset{
		scale only axis,
		xmin=-0.5, xmax=0,
		y axis style/.style={
			yticklabel style=#1,
			ylabel style=#1,
			y axis line style=#1,
			ytick style=#1
		}
	}
\begin{tikzpicture}
	\begin{axis}[xmajorgrids,ymajorgrids,
		scatter/classes={%
			a={blue},%
			b={green},
			c={red}},scatter, no marks,
		scatter src=explicit symbolic,width=0.45\linewidth,
		axis y line*=left,
		y axis style=blue!75!black,
		ymin=-0.2, ymax=0.1,
		xlabel=$x$,
		ylabel=$y$,
		y label style={at={(axis description cs:0.2,.5)}}
		]
		\addplot[scatter,line width=1pt,blue]
		table[meta = class] {
			x        y   class
			-0.0040    0.0025 a
			-0.0081   -0.0038 a
			-0.0121   -0.0091 a
			-0.0162   -0.0051 a
			-0.0202   -0.0026 a
			-0.0243   -0.0090 a
			-0.0283   -0.0143 a
			-0.0323   -0.0103 a 
			-0.0364   -0.0078 a
			-0.0404   -0.0141 a
			-0.0445   -0.0194 a
			-0.0485   -0.0154 a
			-0.0525   -0.0129 a
			-0.0566   -0.0193 a
			-0.0606   -0.0246 a 
			-0.0647   -0.0205 a
			-0.0687   -0.0180 a
			-0.0728   -0.0244 a
			-0.0768   -0.0297 a
			-0.0809   -0.0257 a
			-0.0849   -0.0232 a
			-0.0890   -0.0295 a 
			-0.0930   -0.0348 a
			-0.0970   -0.0308 a
			-0.1011   -0.0283 a
			-0.1051   -0.0347 a
			-0.1091   -0.0400 a
			-0.1132   -0.0360 a
			-0.1172   -0.0335 a
			-0.1213   -0.0398 a
			-0.1253   -0.0451 a
			-0.1294   -0.0411 a
			-0.1334   -0.0386  a
			-0.1375   -0.0449 a
			-0.1415   -0.0502 a
			-0.1456   -0.0462 a
			-0.1496   -0.0437 a
			-0.1537   -0.0501 a
			-0.1577   -0.0554 a
			-0.1617   -0.0514 a
			-0.1658   -0.0489 a
			-0.1698   -0.0552 a
			-0.1738   -0.0605 a
			-0.1779   -0.0565 a
			-0.1819   -0.0540 a
			-0.1860   -0.0603 a
			-0.1900   -0.0656 a
			-0.1941   -0.0616 a
			-0.1981   -0.0591 a
			-0.2022   -0.0655 a
			-0.2062   -0.0708 a
			-0.2103   -0.0668 a
			-0.2143   -0.0643 a
			-0.2184   -0.0706 a
			-0.2224   -0.0759 a
			-0.2264   -0.0719 a
			-0.2305   -0.0694 a
			-0.2345   -0.0758 a
			-0.2385   -0.0811 a
			-0.2426   -0.0771 a
			-0.2466   -0.0745 a
			-0.2507   -0.0809 a
			-0.2547   -0.0862  a
			-0.2588   -0.0822 a
			-0.2628   -0.0797 a
			-0.2669   -0.0860 a
			-0.2709   -0.0913 a
			-0.2750   -0.0873 a
			-0.2790   -0.0848 a
			-0.2831   -0.0912 a
			-0.2871   -0.0965 a
			-0.2911   -0.0925  a
			-0.2952   -0.0900  a
			-0.2992   -0.0963 a
			-0.3032   -0.1016 a
			-0.3073   -0.0976 a
			-0.3113   -0.0951 a
			-0.3154   -0.1014 a
			-0.3194   -0.1067 a
			-0.3235   -0.1027 a
			-0.3275   -0.1002 a
			-0.3316   -0.1066 a
			-0.3356   -0.1119 a
			-0.3397   -0.1079 a
			-0.3437   -0.1054 a
			-0.3477   -0.1117 a
			-0.3518   -0.1170 a
			-0.3558   -0.1130 a
			-0.3599   -0.1105 a
			-0.3639   -0.1169 a
			-0.3679   -0.1222 a
			-0.3720   -0.1181 a
			-0.3760   -0.1156 a
			-0.3801   -0.1220 a
			-0.3841   -0.1273 a
			-0.3882   -0.1233 a
			-0.3922   -0.1208 a
			-0.3963   -0.1271 a
			-0.4003   -0.1324 a
			-0.4044   -0.1284 a
		}; 
	\end{axis}
	\begin{axis}[xmajorgrids,ymajorgrids,
		scatter/classes={%
			a={blue},%
			b={red}},scatter, no marks,
		scatter src=explicit symbolic,width=0.45\linewidth,
		axis y line*=right,
		axis x line=none,
		ymin=-0.2, ymax=0.1,
		y axis style=red!75!black,
		y label style={at={(axis description cs:1.8,.5)}},
		ylabel=$z$
		]
		\addplot[scatter,line width=1pt,red]
		table[meta = class] {
			x        y   class 
			-0.0040   -0.0067 b
			-0.0081   -0.0105 b
			-0.0121   -0.0055 b
			-0.0162    0.0005 b
			-0.0202   -0.0062 b
			-0.0243   -0.0100 b
			-0.0283   -0.0050 b
			-0.0323    0.0011 b
			-0.0364   -0.0056 b
			-0.0404   -0.0095 b
			-0.0445   -0.0044 b
			-0.0485    0.0016 b
			-0.0525   -0.0051 b
			-0.0566   -0.0089 b
			-0.0606   -0.0039 b
			-0.0647    0.0022 b
			-0.0687   -0.0045 b
			-0.0728   -0.0084 b
			-0.0768   -0.0033 b
			-0.0809    0.0027 b
			-0.0849   -0.0040 b
			-0.0890   -0.0078 b
			-0.0930   -0.0028 b
			-0.0970    0.0033 b
			-0.1011   -0.0035 b
			-0.1051   -0.0073 b
			-0.1091   -0.0022 b
			-0.1132    0.0038 b
			-0.1172   -0.0029 b
			-0.1213   -0.0067 b
			-0.1253   -0.0017 b
			-0.1294    0.0044 b
			-0.1334   -0.0024 b
			-0.1375   -0.0062 b
			-0.1415   -0.0011 b
			-0.1456    0.0049 b
			-0.1496   -0.0018 b
			-0.1537   -0.0056 b
			-0.1577   -0.0006 b
			-0.1617    0.0055 b
			-0.1658   -0.0013 b
			-0.1698   -0.0051 b
			-0.1738   -0.0000 b
			-0.1779    0.0060 b
			-0.1819   -0.0007 b
			-0.1860   -0.0045 b
			-0.1900    0.0005 b
			-0.1941    0.0065 b
			-0.1981   -0.0002 b
			-0.2022   -0.0040 b
			-0.2062    0.0010 b
			-0.2103    0.0071 b
			-0.2143    0.0004 b
			-0.2184   -0.0035 b
			-0.2224    0.0016 b
			-0.2264    0.0076 b
			-0.2305    0.0009 b
			-0.2345   -0.0029 b
			-0.2385    0.0021 b
			-0.2426    0.0082 b
			-0.2466    0.0014 b
			-0.2507   -0.0024 b
			-0.2547    0.0027 b
			-0.2588    0.0087 b
			-0.2628    0.0020 b
			-0.2669   -0.0018 b
			-0.2709    0.0032 b
			-0.2750    0.0093 b
			-0.2790    0.0025 b
			-0.2831   -0.0013 b
			-0.2871    0.0038 b
			-0.2911    0.0098 b
			-0.2952    0.0031 b
			-0.2992   -0.0007 b
			-0.3032    0.0043 b
			-0.3073    0.0104 b
			-0.3113    0.0036 b
			-0.3154   -0.0002 b
			-0.3194    0.0049  b
			-0.3235    0.0109 b
			-0.3275    0.0042 b
			-0.3316    0.0004 b
			-0.3356    0.0054 b
			-0.3397    0.0114 b
			-0.3437    0.0047 b
			-0.3477    0.0009 b
			-0.3518    0.0059 b
			-0.3558    0.0120 b
			-0.3599    0.0053 b
			-0.3639    0.0014 b
			-0.3679    0.0065 b
			-0.3720    0.0125 b
			-0.3760    0.0058  b
			-0.3801    0.0020 b
			-0.3841    0.0070 b
			-0.3882    0.0131 b
			-0.3922    0.0064 b
			-0.3963    0.0025 b
			-0.4003    0.0076 b
			-0.4044    0.0136 b
		};
	\end{axis}
\end{tikzpicture}
\end{minipage}
\caption{Trajectory over several periods.}
\end{subfigure}
\begin{subfigure}{.26\linewidth}
\begin{minipage}{0.8\linewidth}
	\centering
	\includegraphics[width=1.2\linewidth]{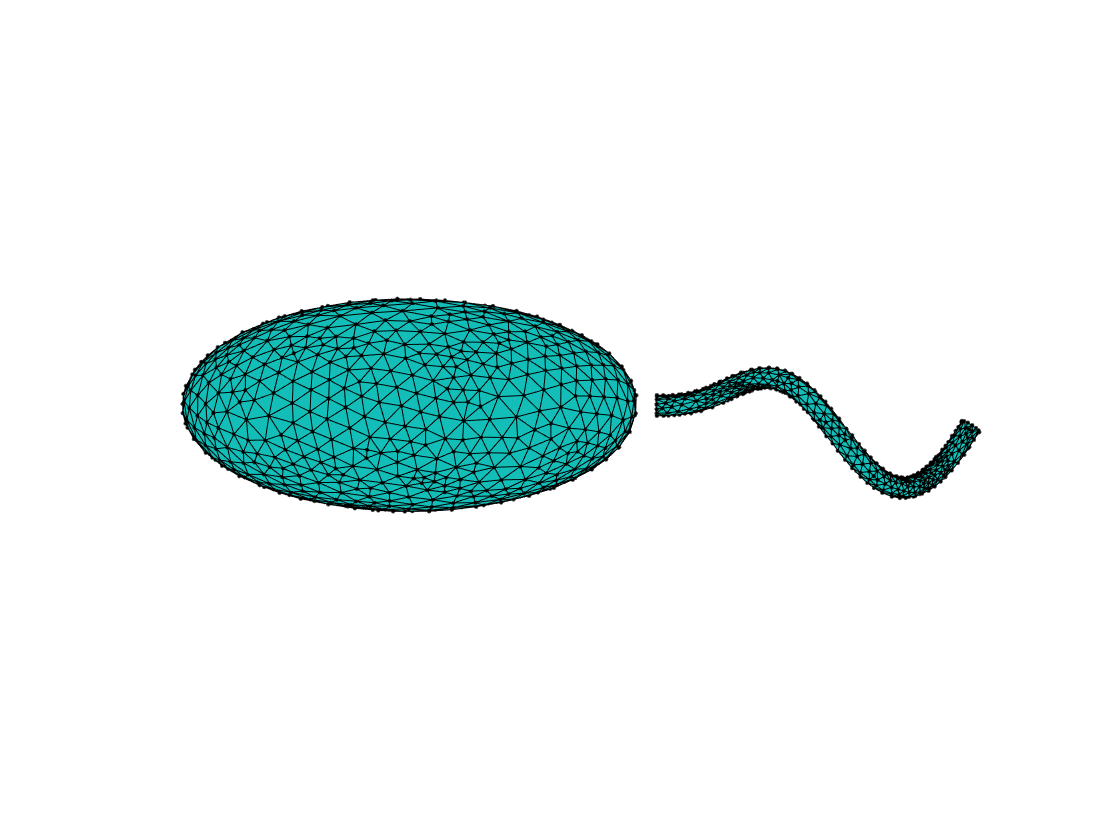}
	\includegraphics[width=1.2\linewidth]{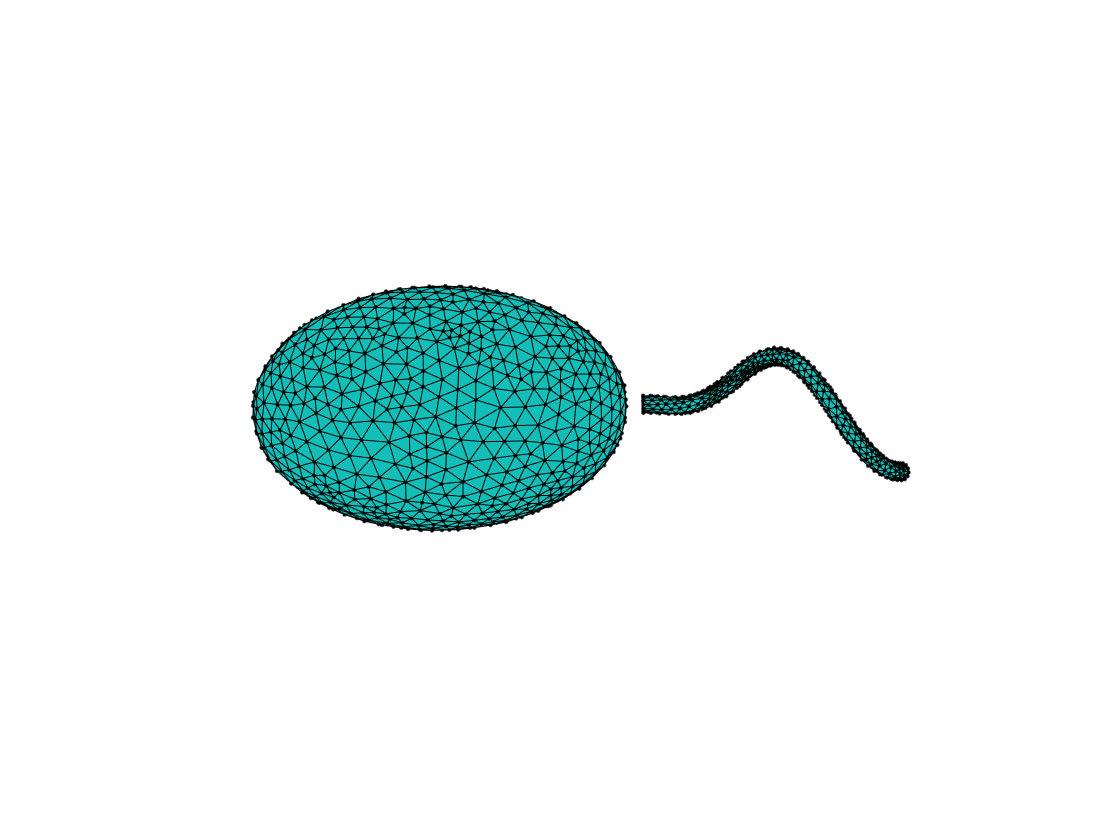}
\end{minipage}
\caption{Top and side view.}
\end{subfigure}
\begin{subfigure}{.7\linewidth}
\begin{minipage}{0.8\linewidth}
	\pgfplotsset{
		scale only axis,
		xmin=-0.5, xmax=0,
		y axis style/.style={
			yticklabel style=#1,
			ylabel style=#1,
			y axis line style=#1,
			ytick style=#1
		}
	}
\begin{tikzpicture}
	\begin{axis}[xmajorgrids,ymajorgrids,
		scatter/classes={%
			a={blue},%
			b={green},
			c={red}},scatter, no marks,
		scatter src=explicit symbolic,width=0.45\linewidth,
		axis y line*=left,
		y axis style=blue!75!black,
		ymin=-0.2, ymax=0.1,
		xlabel=$x$,
		ylabel=$y$,
		y label style={at={(axis description cs:0.2,.5)}},
		]
		\addplot[scatter,line width=1pt,blue]
		table[] {
			x        y  
			-0.0034    0.0033 
			-0.0068    0.0031 
			-0.0103   -0.0004 
			-0.0137   -0.0016 
			-0.0171    0.0017 
			-0.0205    0.0015 
			-0.0239   -0.0020 
			-0.0274   -0.0032 
			-0.0308    0.0001 
			-0.0342   -0.0002 
			-0.0376   -0.0036 
			-0.0410   -0.0049 
			-0.0445   -0.0015 
			-0.0479   -0.0018 
			-0.0513   -0.0052 
			-0.0547   -0.0065 
			-0.0581   -0.0032 
			-0.0616   -0.0034 
			-0.0650   -0.0068 
			-0.0684   -0.0081 
			-0.0718   -0.0048 
			-0.0752   -0.0050 
			-0.0787   -0.0085 
			-0.0821   -0.0097 
			-0.0855   -0.0064 
			-0.0889   -0.0067 
			-0.0923   -0.0101 
			-0.0958   -0.0114 
			-0.0992   -0.0080 
			-0.1026   -0.0083 
			-0.1060   -0.0117 
			-0.1094   -0.0130 
			-0.1128   -0.0097 
			-0.1163   -0.0099 
			-0.1197   -0.0133 
			-0.1231   -0.0146 
			-0.1265   -0.0113 
			-0.1300   -0.0115 
			-0.1334   -0.0150 
			-0.1368   -0.0162 
			-0.1402   -0.0129 
			-0.1436   -0.0132 
			-0.1471   -0.0166 
			-0.1505   -0.0179 
			-0.1539   -0.0145 
			-0.1573   -0.0148 
			-0.1607   -0.0182 
			-0.1642   -0.0195 
			-0.1676   -0.0162 
			-0.1710   -0.0164 
			-0.1744   -0.0198 
			-0.1778   -0.0211 
			-0.1812   -0.0178 
			-0.1847   -0.0180 
			-0.1881   -0.0215 
			-0.1915   -0.0227 
			-0.1949   -0.0194 
			-0.1984   -0.0196 
			-0.2018   -0.0231 
			-0.2052   -0.0244 
			-0.2086   -0.0210 
			-0.2120   -0.0213 
			-0.2155   -0.0247 
			-0.2189   -0.0260 
			-0.2223   -0.0227 
			-0.2257   -0.0229 
			-0.2291   -0.0263 
			-0.2325   -0.0276 
			-0.2360   -0.0243 
			-0.2394   -0.0245 
			-0.2428   -0.0279 
			-0.2462   -0.0292 
			-0.2496   -0.0259 
			-0.2531   -0.0261 
			-0.2565   -0.0296 
			-0.2599   -0.0308 
			-0.2633   -0.0275 
			-0.2667   -0.0278 
			-0.2702   -0.0312 
			-0.2736   -0.0325 
			-0.2770   -0.0291 
			-0.2804   -0.0294 
			-0.2839   -0.0328 
			-0.2873   -0.0341 
			-0.2907   -0.0308 
			-0.2941   -0.0310 
			-0.2975   -0.0344 
			-0.3009   -0.0357 
			-0.3044   -0.0324 
			-0.3078   -0.0326 
			-0.3112   -0.0361 
			-0.3146   -0.0373 
			-0.3180   -0.0340 
			-0.3215   -0.0343 
			-0.3249   -0.0377 
			-0.3283   -0.0390 
			-0.3317   -0.0356 
			-0.3351   -0.0359 
			-0.3386   -0.0393 
			-0.3420   -0.0406 
		};  
	\end{axis}
	\begin{axis}[xmajorgrids,ymajorgrids,
		scatter/classes={%
			a={blue},%
			b={red}},scatter, no marks,
		scatter src=explicit symbolic,width=0.45\linewidth,
		axis y line*=right,
		axis x line=none,
		ymin=-0.2, ymax=0.1,
		y axis style=red!75!black,
		y label style={at={(axis description cs:1.8,.5)}},
		ylabel=$z$,
		]
		\addplot[scatter,line width=1pt,red]
		table[] {
			x        y    
			-0.0034   -0.0009
			-0.0068   -0.0040
			-0.0103   -0.0044
			-0.0137   -0.0016
			-0.0171   -0.0025
			-0.0205   -0.0055
			-0.0239   -0.0060
			-0.0274   -0.0031
			-0.0308   -0.0040
			-0.0342   -0.0071
			-0.0376   -0.0076
			-0.0410   -0.0047
			-0.0445   -0.0056
			-0.0479   -0.0087
			-0.0513   -0.0091
			-0.0547   -0.0063
			-0.0581   -0.0072
			-0.0616   -0.0102
			-0.0650   -0.0107
			-0.0684   -0.0079
			-0.0718   -0.0087
			-0.0752   -0.0118
			-0.0787   -0.0123
			-0.0821   -0.0094
			-0.0855   -0.0103
			-0.0889   -0.0134
			-0.0923   -0.0139
			-0.0958   -0.0110
			-0.0992   -0.0119
			-0.1026   -0.0150
			-0.1060   -0.0154
			-0.1094   -0.0126
			-0.1128   -0.0135
			-0.1163   -0.0165
			-0.1197   -0.0170
			-0.1231   -0.0141
			-0.1265   -0.0150
			-0.1300   -0.0181
			-0.1334   -0.0186
			-0.1368   -0.0157
			-0.1402   -0.0166
			-0.1436   -0.0197
			-0.1471   -0.0201
			-0.1505   -0.0173
			-0.1539   -0.0182
			-0.1573   -0.0212
			-0.1607   -0.0217
			-0.1642   -0.0189
			-0.1676   -0.0197
			-0.1710   -0.0228
			-0.1744   -0.0233
			-0.1778   -0.0204
			-0.1812   -0.0213
			-0.1847   -0.0244
			-0.1881   -0.0249
			-0.1915   -0.0220
			-0.1949   -0.0229
			-0.1984   -0.0260
			-0.2018   -0.0264
			-0.2052   -0.0236
			-0.2086   -0.0245
			-0.2120   -0.0275
			-0.2155   -0.0280
			-0.2189   -0.0251
			-0.2223   -0.0260
			-0.2257   -0.0291
			-0.2291   -0.0296
			-0.2325   -0.0267
			-0.2360   -0.0276
			-0.2394   -0.0307
			-0.2428   -0.0312
			-0.2462   -0.0283
			-0.2496   -0.0292
			-0.2531   -0.0322
			-0.2565   -0.0327
			-0.2599   -0.0299
			-0.2633   -0.0307
			-0.2667   -0.0338
			-0.2702   -0.0343
			-0.2736   -0.0314
			-0.2770   -0.0323
			-0.2804   -0.0354
			-0.2839   -0.0359
			-0.2873   -0.0330
			-0.2907   -0.0339
			-0.2941   -0.0370
			-0.2975   -0.0374
			-0.3009   -0.0346
			-0.3044   -0.0355
			-0.3078   -0.0385
			-0.3112   -0.0390
			-0.3146   -0.0361
			-0.3180   -0.0370
			-0.3215   -0.0401
			-0.3249   -0.0406
			-0.3283   -0.0377
			-0.3317   -0.0386
			-0.3351   -0.0417
			-0.3386   -0.0422
			-0.3420   -0.0393
		};
\end{axis}
\end{tikzpicture}
\end{minipage}
\caption{Trajectory over several periods.}
\end{subfigure}
\begin{subfigure}{0.26\linewidth}
\begin{minipage}{0.8\linewidth}
	\includegraphics[width=\linewidth]{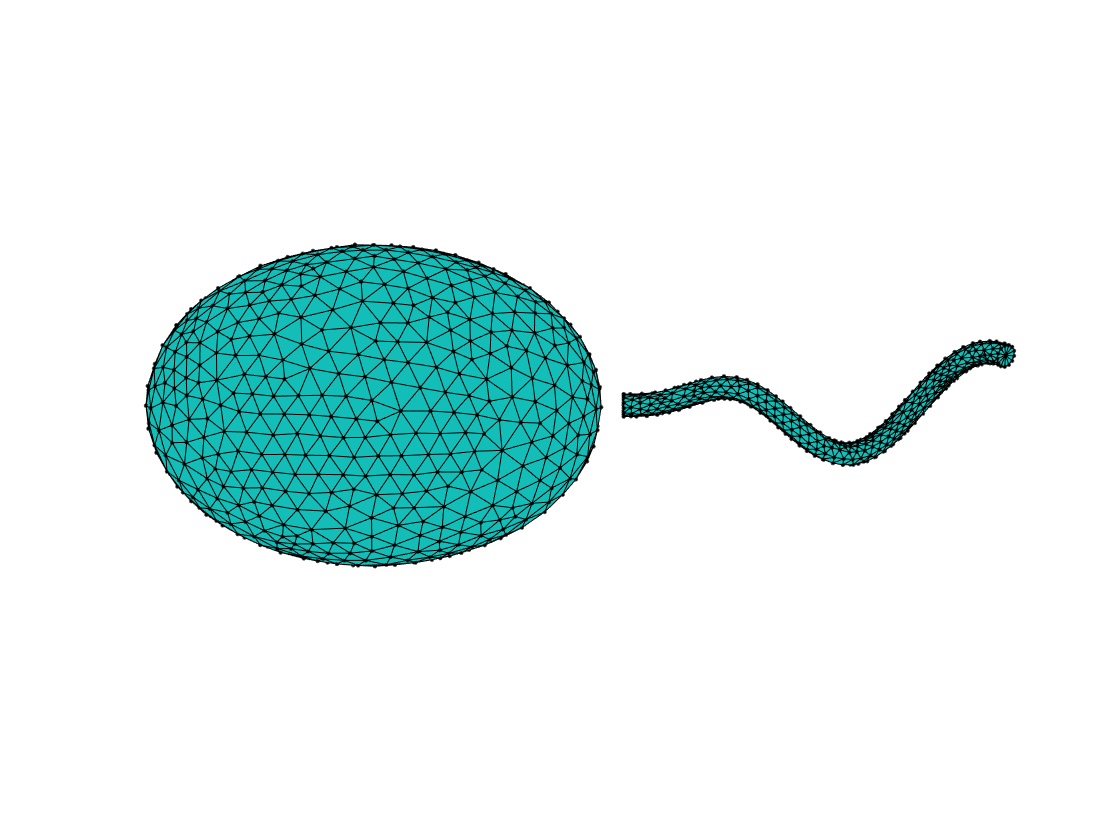}
	\includegraphics[width=\linewidth]{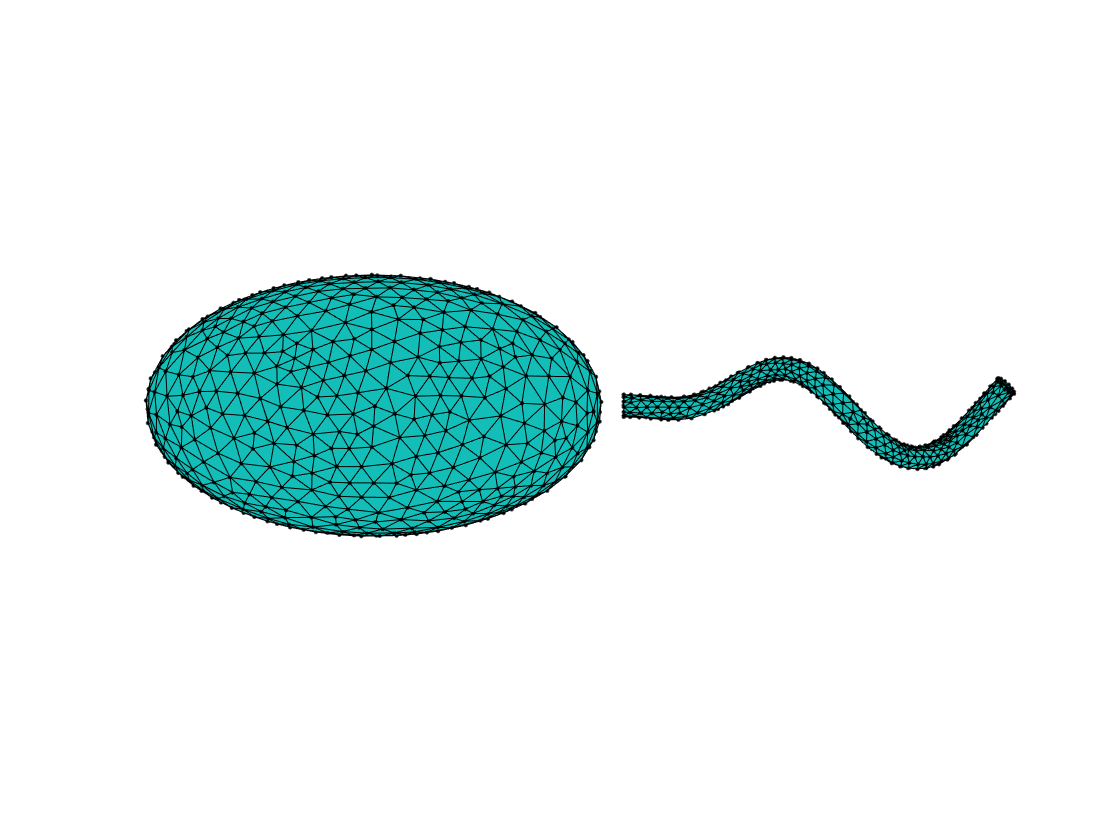}
\end{minipage}
\caption{Top and side view.}
\end{subfigure}
\begin{subfigure}{0.7\linewidth}
\begin{minipage}{\linewidth}
	\pgfplotsset{
	scale only axis,
	xmin=-0.5, xmax=0,
	y axis style/.style={
		yticklabel style=#1,
		ylabel style=#1,
		y axis line style=#1,
		ytick style=#1
	}
}
\begin{tikzpicture}
\begin{axis}[xmajorgrids,ymajorgrids,
scatter/classes={%
	a={blue},%
	b={green},
	c={red}},scatter, no marks,
scatter src=explicit symbolic,width=0.45\linewidth,
axis y line*=left,
y axis style=blue!75!black,
ymin=-0.2, ymax=0.1,
y label style={at={(axis description cs:0.2,.5)}},
xlabel=$x$,
ylabel=$y$
]
\addplot[scatter,line width=1pt,blue]
table[] {
	x        y  
 -0.0024    0.0001
-0.0048    0.0004
-0.0072    0.0005
-0.0097    0.0002
-0.0121    0.0003
-0.0145    0.0006
-0.0169    0.0006
-0.0193    0.0003
-0.0217    0.0004
-0.0242    0.0007
-0.0266    0.0008
-0.0290    0.0005
-0.0314    0.0006
-0.0338    0.0009
-0.0362    0.0010
-0.0387    0.0007
-0.0411    0.0008
-0.0435    0.0011
-0.0459    0.0011
-0.0483    0.0009
-0.0507    0.0009
-0.0532    0.0013
-0.0556    0.0013
-0.0580    0.0010
-0.0604    0.0011
-0.0628    0.0014
-0.0652    0.0015
-0.0677    0.0012
-0.0701    0.0013
-0.0725    0.0016
-0.0749    0.0017
-0.0773    0.0014
-0.0797    0.0014
-0.0822    0.0018
-0.0846    0.0018
-0.0870    0.0015
-0.0894    0.0016
-0.0918    0.0019
-0.0942    0.0020
-0.0966    0.0017
-0.0991    0.0018
-0.1015    0.0021
-0.1039    0.0022
-0.1063    0.0019
-0.1087    0.0020
-0.1111    0.0023
-0.1136    0.0023
-0.1160    0.0021
-0.1184    0.0021
-0.1208    0.0025
-0.1232    0.0025
-0.1256    0.0022
-0.1281    0.0023
-0.1305    0.0026
-0.1329    0.0027
-0.1353    0.0024
-0.1377    0.0025
-0.1401    0.0028
-0.1426    0.0029
-0.1450    0.0026
-0.1474    0.0026
-0.1498    0.0030
-0.1522    0.0030
-0.1546    0.0027
-0.1571    0.0028
-0.1595    0.0031
-0.1619    0.0032
-0.1643    0.0029
-0.1667    0.0030
-0.1691    0.0033
-0.1715    0.0034
-0.1740    0.0031
-0.1764    0.0032
-0.1788    0.0035
-0.1812    0.0035
-0.1836    0.0033
-0.1860    0.0033
-0.1885    0.0037
-0.1909    0.0037
-0.1933    0.0034
-0.1957    0.0035
-0.1981    0.0038
-0.2005    0.0039
-0.2030    0.0036
-0.2054    0.0037
-0.2078    0.0040
-0.2102    0.0041
-0.2126    0.0038
-0.2150    0.0038
-0.2175    0.0042
-0.2199    0.0042
-0.2223    0.0039
-0.2247    0.0040
-0.2271    0.0043
-0.2295    0.0044
-0.2320    0.0041
-0.2344    0.0042
-0.2368    0.0045
-0.2392    0.0046
-0.2416    0.0043
};  
\end{axis}
\begin{axis}[xmajorgrids,ymajorgrids,
scatter/classes={%
	a={blue},%
	b={red}},scatter, no marks,
scatter src=explicit symbolic,width=0.45\linewidth,
axis y line*=right,
axis x line=none,
ymin=-0.2, ymax=0.1,
y label style={at={(axis description cs:1.8,.5)}},
ylabel=$z$,
y axis style=red!75!black
]
\addplot[scatter,line width=1pt,red]
table[] {
	x        y    
 -0.0024    0.0003
-0.0048    0.0004
-0.0072    0.0000
-0.0097   -0.0001
-0.0121    0.0002
-0.0145    0.0002
-0.0169   -0.0001
-0.0193   -0.0003
-0.0217    0.0001
-0.0242    0.0001
-0.0266   -0.0003
-0.0290   -0.0004
-0.0314   -0.0001
-0.0338   -0.0001
-0.0362   -0.0004
-0.0387   -0.0006
-0.0411   -0.0002
-0.0435   -0.0002
-0.0459   -0.0006
-0.0483   -0.0007
-0.0507   -0.0004
-0.0532   -0.0004
-0.0556   -0.0007
-0.0580   -0.0009
-0.0604   -0.0005
-0.0628   -0.0005
-0.0652   -0.0009
-0.0677   -0.0010
-0.0701   -0.0007
-0.0725   -0.0007
-0.0749   -0.0010
-0.0773   -0.0012
-0.0797   -0.0008
-0.0822   -0.0008
-0.0846   -0.0012
-0.0870   -0.0013
-0.0894   -0.0010
-0.0918   -0.0010
-0.0942   -0.0013
-0.0966   -0.0015
-0.0991   -0.0011
-0.1015   -0.0011
-0.1039   -0.0015
-0.1063   -0.0016
-0.1087   -0.0013
-0.1111   -0.0013
-0.1136   -0.0016
-0.1160   -0.0018
-0.1184   -0.0014
-0.1208   -0.0014
-0.1232   -0.0018
-0.1256   -0.0019
-0.1281   -0.0016
-0.1305   -0.0016
-0.1329   -0.0019
-0.1353   -0.0021
-0.1377   -0.0017
-0.1401   -0.0017
-0.1426   -0.0020
-0.1450   -0.0022
-0.1474   -0.0019
-0.1498   -0.0019
-0.1522   -0.0022
-0.1546   -0.0024
-0.1571   -0.0020
-0.1595   -0.0020
-0.1619   -0.0023
-0.1643   -0.0025
-0.1667   -0.0022
-0.1691   -0.0021
-0.1715   -0.0025
-0.1740   -0.0026
-0.1764   -0.0023
-0.1788   -0.0023
-0.1812   -0.0026
-0.1836   -0.0028
-0.1860   -0.0025
-0.1885   -0.0024
-0.1909   -0.0028
-0.1933   -0.0029
-0.1957   -0.0026
-0.1981   -0.0026
-0.2005   -0.0029
-0.2030   -0.0031
-0.2054   -0.0027
-0.2078   -0.0027
-0.2102   -0.0031
-0.2126   -0.0032
-0.2150   -0.0029
-0.2175   -0.0029
-0.2199   -0.0032
-0.2223   -0.0034
-0.2247   -0.0030
-0.2271   -0.0030
-0.2295   -0.0034
-0.2320   -0.0035
-0.2344   -0.0032
-0.2368   -0.0032
-0.2392   -0.0035
-0.2416   -0.0037
};
\end{axis}
\end{tikzpicture}
\end{minipage}
\caption{Trajectory over several periods.}
\end{subfigure}
\caption{Optimal swimmer in the monoflagellated case, with different constraints on the transversal velocities $|\bar{U}_y|, |\bar{U}_z|$. In subfigure (a)-(b) $|\bar{U}_y|, |\bar{U}_z|\le 0.1$, (c)-(d) $|\bar{U}_y|, |\bar{U}_z| \le 0.01$, (e)-(f) $|\bar{U}_y|, |\bar{U}_z| \le 0.001$.}
\label{Fig: OptiHead1b}
\end{figure}

Different values of the latter velocity constraints are tested, and their effects are visible on the swimmers' optimal shapes. In figure \ref{Fig: OptiHead1b}a, the least constrained swimmer is presented: in this case, the lateral velocities are of the same order of magnitude of the horizontal one. We observe that the optimal tail reaches the maximal possible value of $R^t$, while the optimal head has a prolate shape with $R_1^h$ close to the maximal value 1.5. Having large $R^t$ and $\lambda$ implies that the tail remains funnel-shaped and asymmetric, ensuring larger propulsion force. At the same time large $R_1^h$ (and consequently smaller $R_2^h$, $R_3^h$) reduces the drag resistance of the ellipsoid to an incident flow parallel to its horizontal axis.
The mildly constrained optimal swimmer is depicted in figure \ref{Fig: OptiHead1b}c. Only $R_1^h$ is close to the upper bound of its feasible region. Lower values of $R^t$ and $\lambda$ entail lower propulsion forces, and thus lower speed, ensuring a propulsion mainly in the horizontal direction. Analogously, lower values of $R_1^h$ and larger values of $R_2^h, R_3^h$ decrease the hydrodynamicity of the head but they increase its stability. The resulting shape is still of prolate type. The smaller values of the tail parameters increase its symmetry with respect to the propulsion axis. Further tightening the constraints produces an even more symmetrical tail, hence an even slower but much more stable swimmer with a less elongated head, as shown in figure \ref{Fig: OptiHead1b}e. 
Figure \ref{Fig: Obj1tail} shows that the performance of the swimmer improves mainly in the first 30 iterations. The figure just presents the objective function of the mildly constrained case, associated with figure \ref{Fig: OptiHead1b}c-d, as the two other constrained cases present similar behaviours.

The numerical values of the optimal shape parameters are collected in Table \ref{Table:Optivalues1}.
\begin{center}
	\begin{table}[h]
		\begin{tabular}{c  c  c  c c c}
			\hline 
			&$R^t$ & $\lambda$&$R_1^h$ &$R_2^h$ &$R_3^h$ \\
			Figure \ref{Fig: OptiHead1b}a&1.0     &  2.58   &     1.49 &    0.89 & 0.75\\
			Figure \ref{Fig: OptiHead1b}c&	0.58  &     2.12  &   1.47   &  1.0 & 0.68\\
			Figure \ref{Fig: OptiHead1b}e&0.32   &   1.73   &   1.35   &  0.78& 0.95\\
			\hline
		\end{tabular}
		\caption{Parameters describing the geometry of the monoflagellated optimal swimmers in figure \ref{Fig: OptiHead1b}.}
		\label{Table:Optivalues1}
	\end{table}
\end{center}
\subsection{Biflagellated swimmer}
\begin{figure}
	
	\resizebox{0.4\textwidth}{!}{\begin{tikzpicture}
		
		\draw[ultra thick]  (-4,0) ellipse (2.5cm and \radius);
		\draw[thick]  (-8,0) -- (0,0);
		
		\draw[thick]  (-4,0) -- (-1.6,0.5);
		\draw[ultra thick, blue] (-4,0) -- (-2,0.9);
		\node[blue] at (-0.8,0.5) { \huge $\mathbf{\alpha}$};
		\draw[thick]  (-4,0) -- (-3.5,1.4);
		
		\draw[thick] (-2,0.9) -- (-1.6,0.7);
		\draw[ultra thick, red] (-2,0.9) -- (-1.6,1.1);
		\node[red] at (-1.4,1.1) { \huge $\mathbf{\gamma}$};
		\draw[thick]  (-2,0.9) -- (-1.6,0.9);
		
		\draw[thick]  (-3.5,1.4) -- (-3,1.9);
		\draw[thick]  (-3.5,1.4) -- (-3,1.5);
		\draw[thick]  (-3.5,1.4) -- (-3,1.7);
		
		\draw[thick]  (-1.6,0.5) -- (-1.2,0.5);
		\draw[thick]  (-1.6,0.5) -- (-1.2,0.7);
		
		\draw[->,ultra thick,black] (-2,1.1)   to [out=45,in=135] (0.5,1.1) ;
		
		\node[inner sep=0pt] at (5,0)
		{\includegraphics[width=0.5\textwidth]{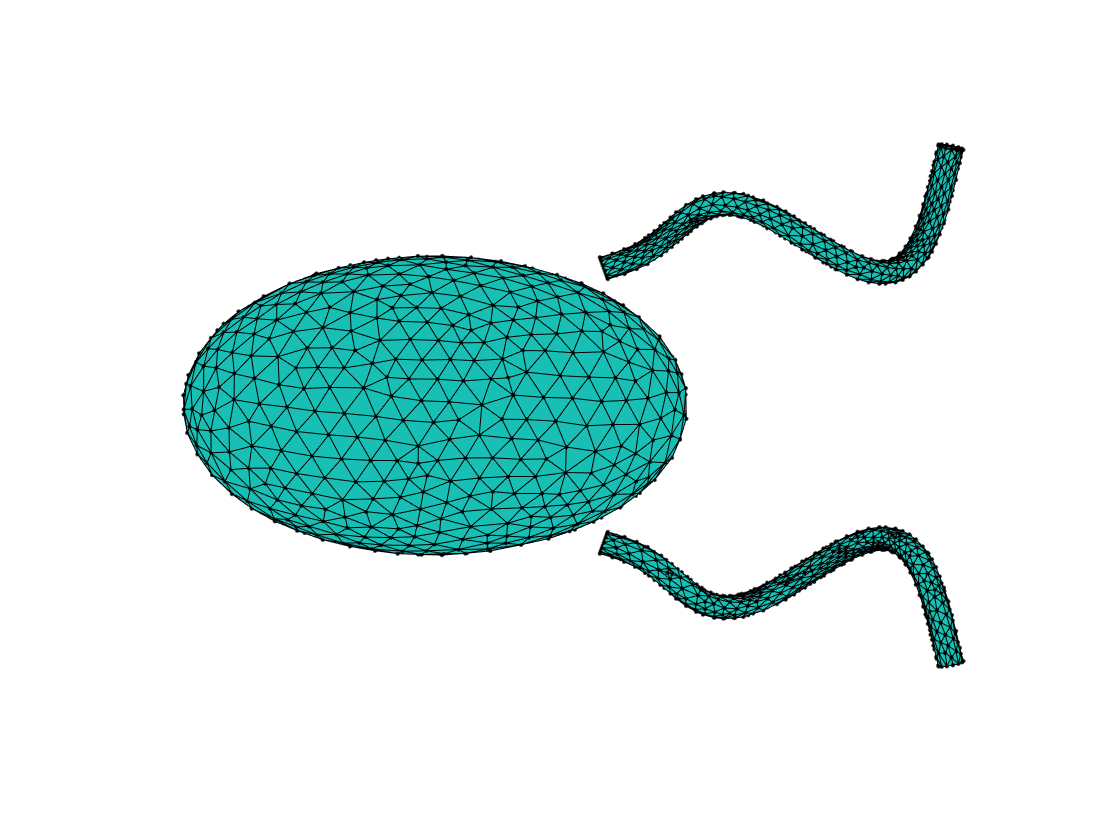}};
		\end{tikzpicture}
	}
	\caption{Schematization of $(\alpha,\gamma)$ variability. In the figure, an example is reproduced where $(\alpha,\gamma) = (0.3\pi,0.1\pi)$.}
	\label{Fig:Schematization-tails}
\end{figure}

The presence of two flagella restores the symmetry of the swimmer's shape, which was an issue in the monoflagellated case. In addition to the tail and cell body parameters, in this case it is possible to vary the position of the tails and their inclination while maintaining the symmetric shape of the swimmer.
The placement of the helices is parametrized by two angles: the angle $\alpha$ describes the latitude of the tail junction, while the angle $\gamma$ describes the inclination of the tail's axis with respect to the  horizontal plane. Figure \ref{Fig:Schematization-tails} gives a graphical representation of the roles of $\alpha$ and $\gamma$ in the position of the tail junctions.

\begin{figure}[h]
	\begin{tikzpicture}
\node[anchor=south west,inner sep=0] (image) at (0,0){
	\includegraphics[width=0.45\linewidth]{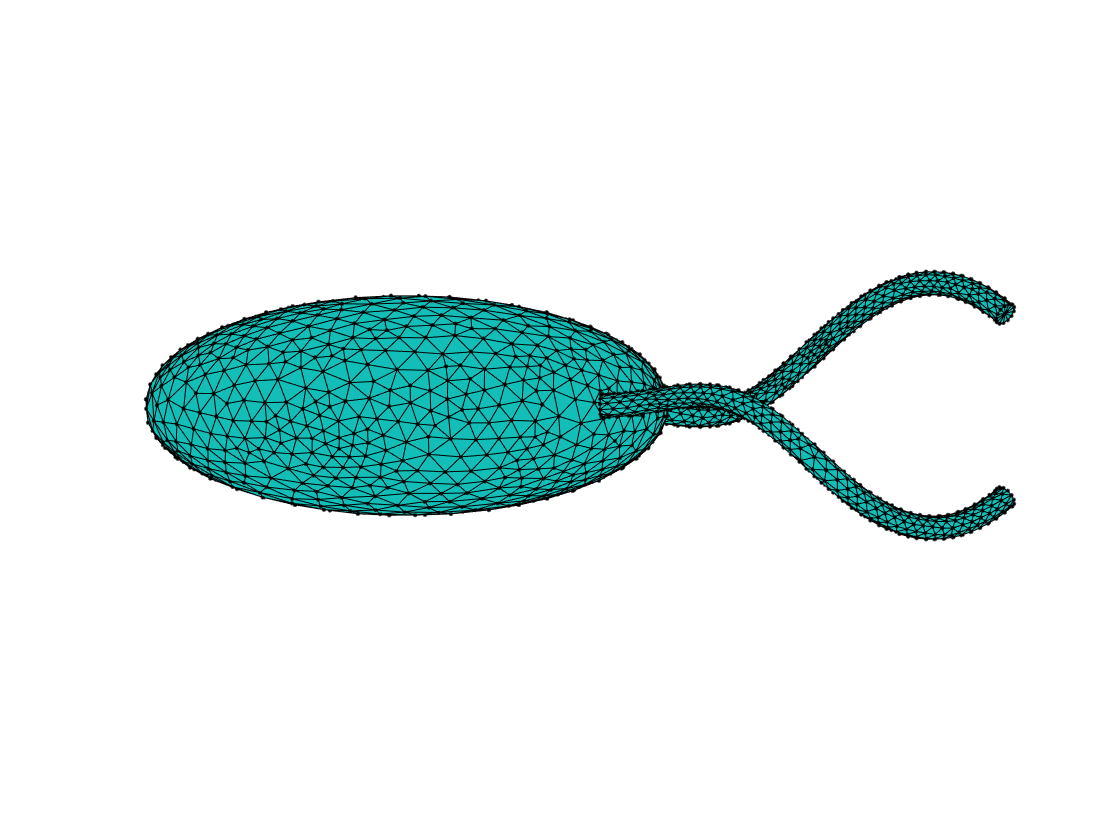}
	\includegraphics[width=0.45\linewidth]{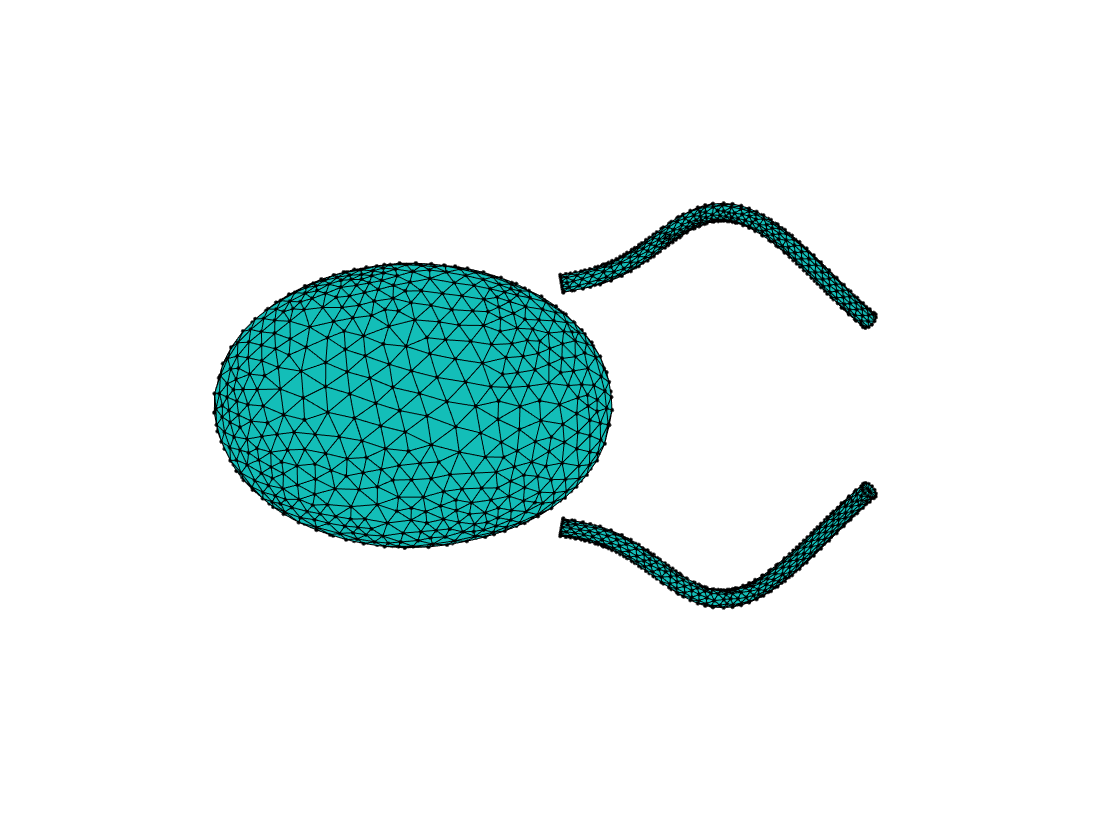}};
\begin{scope}[x={(image.south east)},y={(image.north west)}]
\node at (0.22,0) [ align=left]{Top view.};
\node at (0.72,0) [ align=left]{Side view.};
\end{scope}
\end{tikzpicture}
	\caption{Optimal swimmer in the biflagellated case.}
	\label{Fig: OptiHead2}
\end{figure}

The optimal swimmer is depicted in figure \ref{Fig: OptiHead2}, and it has a prolate head, i.e. elongated in the propulsion direction, as the monoflagellated swimmer. The ellipsoid is flattened in the perpendicular direction to the plane that contains the tails' junctions, $R_2^h$, while the radius $R_1^h$ attains its upper bound. Since the symmetry is ensured by the overall body-tail configuration, it is no longer necessary to have a symmetric tail shape to propel along a straight line. For this reason, the optimal tails only need to focus on propulsion and not on the stability, which explains the large values of the helix radius $R^t$ and its wavelength $\lambda$. The optimal angular parameters are $\alpha \approx 0.27 \pi$ and $\gamma \approx 0.06\pi$, which correspond to closer tails but slightly tilted far from each other. The optimal parameters are presented in Table \ref{Table:Optivalues2}.

In this case, the optimization algorithm was launched with a set of 50 shapes uniformly sampled in the feasible space, for which the objective function was evaluated. Additional twenty evaluations are performed, and figure \ref{Fig: Obj2tail} shows that only an  improvement of order 2\% was produced with respect to the best input. 

\paragraph*{Remark.}When dealing with multiple tails, it is possible that the optimization algorithm proposes sets of parameters that lead to tail-tail intersection or tail-body intersection during its exploration of the parameter space. These sets of parameters are declared unfeasible, whenever they appear.
\begin{center}
	\begin{table}
		\begin{tabular}{c  c  c c c c c}
			\hline 
			$R^t$ & $\lambda$&$R_1^h$ &$R_2^h$ &$R_3^h$&$\alpha$&$\gamma$ \\
			1.0  &    3.46  &     1.49  &   0.63 & 1.07&   0.85&    0.17 \\
			\hline
		\end{tabular}
		\caption{Parameters describing the geometry of the optimal bacterium in figure \ref{Fig: OptiHead2}.}
		\label{Table:Optivalues2}
	\end{table}
\end{center}

\subsection{Tetra-flagellated swimmer}
The four flagella are located at an angular distance of $\frac{\pi}{2}$ from each other, ensuring a symmetric body configuration (see figure \ref{Fig:positionTail}) which prevents transversal displacements.
In the rest we fix $\alpha=0.45\pi$ and $\gamma=0$.

The optimal shape of the tetra-flagellated swimmer is depicted in figure \ref{Fig: OptiHead4}. The resulting tails have a funnel-like shape, close to the monoflagellated least constrained case (see Figure \ref{Fig: OptiHead1b}a) and the biflagellated case (see Figure \ref{Fig: OptiHead2}). Surprisingly, regarding the head,  the optimal shape is almost spherical. Table \ref{Table:Optivalues4} contains the values of the optimal parameters describing the tetra-flagellated swimmer.

The algorithm converges in almost 40 iterations (see Figure \ref{Fig: Obj4tail}) and the result given by the first 10 iterations differs by 9\% from the final result.
\begin{figure}
		\resizebox{0.4\textwidth}{!}{\begin{tikzpicture}
		\draw[ultra thick]  (-5,0) ellipse (2.5cm and \radius);
		
		\draw[ultra thick] (-5.5,-2.4) -- (-5.5,1.8) -- (-4.6,2.4) -- (-4.6,-1.8) -- (-5.5,-2.4);		
		
		\draw[ thick,decoration={aspect=0.31, segment length=1mm,
			amplitude=0.1cm,coil},decorate] (-3.5,\radius+3*\smallradius) -- (-5,\radius+3*\smallradius);
		
		\draw[ thick,decoration={aspect=0.31, segment length=1mm,
			amplitude=0.1cm,coil},decorate] (-3.5,-\radius-3*\smallradius) -- (-5,-\radius-3*\smallradius);
		
		\draw[thick,decoration={aspect=0.31, segment length=1mm,
			amplitude=0.1cm,coil},decorate] (-3.9,0) -- (-5.4,0);
		
		\draw[ultra thick]  (0,0) circle (\radius);
		\draw[ultra thick] (0,\radius+3*\smallradius) circle (\smallradius);
		\draw[ultra thick]  (\radius+3*\smallradius,0) circle (\smallradius);
		\draw[ultra thick]  (-\radius-3*\smallradius,0) circle (\smallradius);
		\draw[ultra thick]  (0,-\radius-3*\smallradius) circle (\smallradius);
		
		\draw[ thick] (-1.5*\smallradius,\radius+3*\smallradius) -- (1.5*\smallradius,\radius+3*\smallradius) ;
		\node at (0,\radius+6*\smallradius) [ align=center]{2r};
		\draw[thick] (0*\smallradius,-\radius+0.5*\smallradius) -- (0*\smallradius,-\radius-2.5*\smallradius) ;
		\node at (-1.5*\smallradius,-\radius-1.5*\smallradius) [ align=center]{$l$};
		\node at (0,0*\radius) [ align=left]{\large Head};
		\node at (3*\radius,-0.5*\radius)  [ align=left]{\large Tails};
		
		\draw[->]   (2.5*\radius,-0.5*\radius) -- (2*\smallradius,\radius+3*\smallradius) ;
		\draw[->]   (2.5*\radius,-0.5*\radius) -- (2*\smallradius,-\radius-3*\smallradius) ;
		\draw[->]   (2.5*\radius,-0.5*\radius) -- (\radius+5*\smallradius,0 ) ;
		\draw[->]   (2.5*\radius,-0.5*\radius) -- (-\radius-\smallradius,0 ) ;
		\end{tikzpicture}}
\caption{Position of the tails for the tetra-flagellated swimmer. Lateral and frontal view.}
\label{Fig:positionTail}
\end{figure}
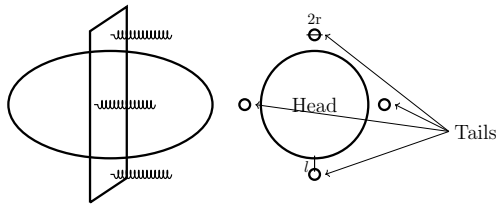
\begin{figure}[h]
	\begin{tikzpicture}
	\node[anchor=south west,inner sep=0] (image) at (0,0){
		\includegraphics[width=0.45\linewidth]{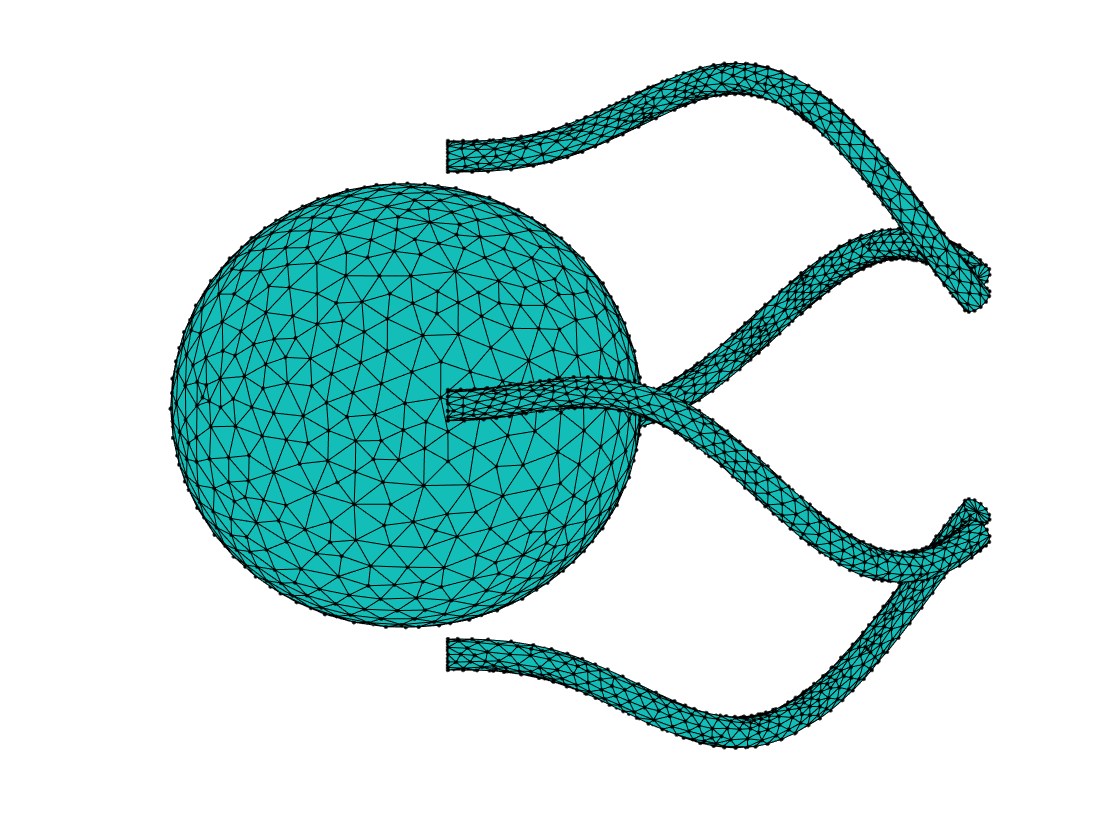}
		\includegraphics[width=0.45\linewidth]{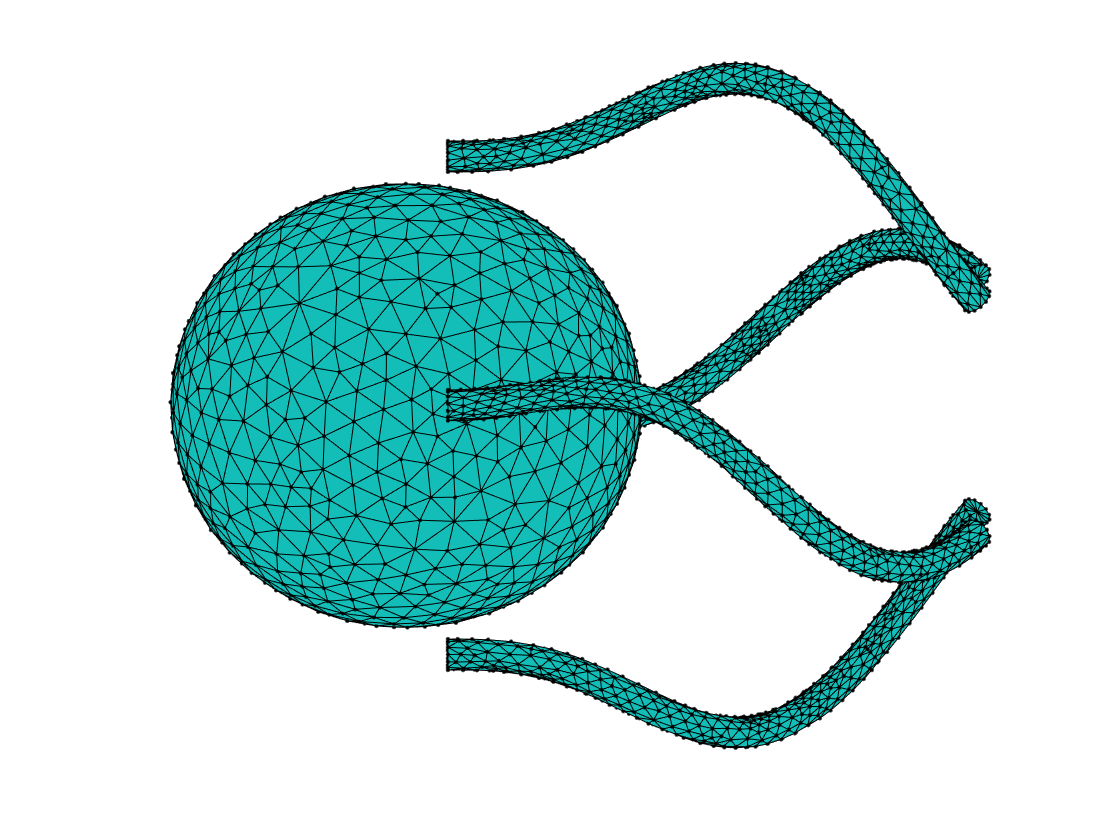}};
	\begin{scope}[x={(image.south east)},y={(image.north west)}]
	\node at (0.22,0) [ align=left]{Top view.};
	\node at (0.72,0) [ align=left]{Side view.};
	\end{scope}
	\end{tikzpicture}
	\caption{Optimal swimmer in the tetra-flagellated case.}
	\label{Fig: OptiHead4}
\end{figure}
\begin{center}
	\begin{table}
		\begin{tabular}{c  c  c c c}
			\hline 
			$R^t$ & $\lambda$&$R_1^h$ &$R_2^h$ &$R_3^h$ \\
			 1.0   &    3.59   &   1.01  &  0.96 & 1.03\\
			\hline
		\end{tabular}
		\caption{Parameters describing the geometry of the optimal bacterium in figure \ref{Fig: OptiHead4}.}
		\label{Table:Optivalues4}
	\end{table}
\end{center}
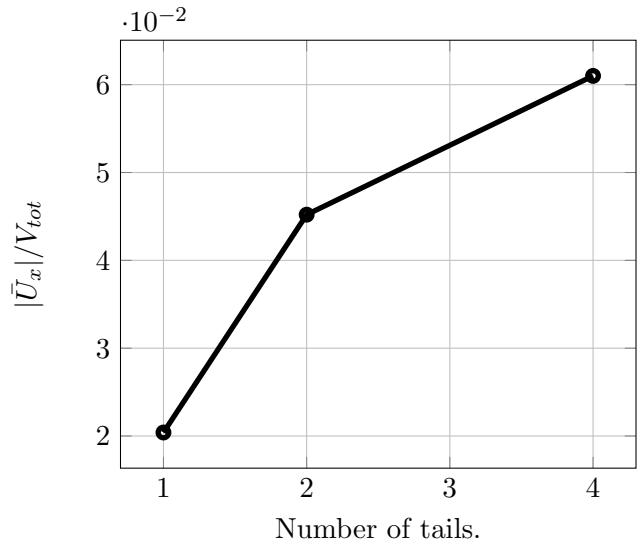
\begin{figure}
\begin{tikzpicture}
	\begin{axis}[xlabel={$\text{Number of tails.}$}, ylabel={$|\bar{U}_x|/V_{tot}$},xmajorgrids,ymajorgrids,
	xtick={1,2,3,4,5},
	scatter/classes={%
		b={mark=o,black}},
	scatter src=explicit symbolic]
	\addplot[scatter,line width=2pt,black]
	table[meta = class] {
x        y   class
1    0.0204 b 
2  0.0452 b
4    0.0610 b
	};
	\end{axis}
	\end{tikzpicture}
	\caption{Speed comparison between the fastest monoflagellated, biflagellated and tetra-flagellated microswimmers. The mean propulsion speed is normalized by the total volume of the swimmer. We notice that passing from 2 to 4 does not produce a doubling of the advancement speed as passing from 1 to 2 produces. The flagella are slender and their volume is $1\%$ of the cell volume.}
	\label{Fig: Speed comparison}
\end{figure}

\section{Discussion}
\label{Sect:Discussion}
Figure \ref{Fig: Speed comparison} compares the speeds of the optimal mildly constrained monoflagellated, biflagellated and tetra-flagellated swimmers. 

Adding a flagellum to a monoflagellated swimmer almost doubles the propulsion speed, while the tetra-flagellated case produces a less significant increase with respect to the biflagellated one. This is due to the fact that, the more the flagella, the stronger the mutual interaction which inhibits the propulsive potential of multiple tails.
This behaviour was already observed in experimental studies \cite{Ye2013}.
Moreover, our results show that funnel-like tails produce larger propulsion speeds. 

The numerical experiments that were conducted show that the optimal shape of a multi-flagellated swimmer depends strongly on the number of flagella and their position. 
In the case of a monoflagellated or biflagellated swimmer, the body shapes we found were elongated in the direction of the motion, while for a tetra-flagellated swimmer a spherical cell body is preferred. 
Our results indicate that the position and number of flagella modify the propulsion pattern and play a significant role in the optimal design of the head.
This argument also justifies the multiple natural head shape of helical bacteria. Indeed, for instance, Escherichia coli bacteria have elongated head while mediterranean magneto-ovoid bacterium MO-1 have a rounder shape.
\section{Perspectives}
Further investigations could be conducted by taking into account the elasticity at the tails' junctions or elastic deformable tails. 
Also, generalizing the geometrical shapes considered would be an other perspectives.
Bayesian optimization could be applied in a more complex framework, as for swimmers immersed into a non-Newtonian fluid.
\section*{Acknowledgments}
L. Berti is funded by Labex IRMIA.
%

\appendix
\section{Explicit formulas for matrices in \eqref{Eq:Swimming-disc}}
\label{Appendix}
The block form of equations \eqref{Eq:Swimming-disc} is
\begin{equation*}
\begin{tabular}{r}
\scalebox{0.75}{$3\times N$} $\left\{\lefteqn{\phantom{\begin{matrix} G \end{matrix}}}\right.$\\
\scalebox{0.75}{$3$} $\left\{\lefteqn{\phantom{\begin{matrix} J \end{matrix}}} \right.$\\
\scalebox{0.75}{$3$} $\left\{\lefteqn{\phantom{\begin{matrix} 0 \end{matrix}}} \right.$
\end{tabular}
\begin{bmatrix}
\, \, G & J^T & K^T \\
\, \,  J & 0 & 0 \\
\, \, \, \, \, \, \myunderbrace{K}{3\times N} & \myunderbrace{0}{3} & \myunderbrace{0}{3} 
\end{bmatrix}
\begin{bmatrix}
f \\
U \\
\Omega
\end{bmatrix}
= 
\begin{bmatrix}
I(\omega)\\
0 \\
0
\end{bmatrix}.
\end{equation*}

\medskip
Submatrix $G$ is composed of $(1+n_T)^2$ submatrices $G^{\{A,B\}}$, for $A,B \in \{\partial H, \partial F_i\}$. Define as $N_A$ and $N_B$ the cardinality of the scalar finite element subspaces corresponding to the degrees of freedom over $A$ and $B$, respectively. Each of the $G^{\{A,B\}}$ is subdivided into $N_A\times N_B$ submatrices of size $3\times3$, named $G_{ij,lk}^{\{A,B\}}$ defined as
\begin{equation}
	G_{ij,lk}^{\{A,B\}} := \int_A\int_B G_{ij}(x,y)\phi_l(x)\phi_k(y) \, \mathrm{d}x  \mathrm{d}y, 
\end{equation}
for $i,j=1,2,3$ and  $l=1,\dots,N_B$, $k=1,\dots,N_A$.

Submatrix $J$ is composed of $(1+n_T)$ submatrices $J^{A}$ for $A \in \{\partial H, \partial F_i\}$. Each of the $J^{A}$ has size $3\times 3N_A$, it is block diagonal and its components $J^{A}_{il,j}$ are $3\times N_A$ submatrices defined as
\begin{equation}
	J^{A}_{il,j} :=\int_A \vec{e}_i\phi_l(x) \, \mathrm{d}x,
\end{equation}\
for $i=1,2,3$ and $l=1,\dots,N_A$ inside each block and $j=1,2,3$ denoting the diagonal block.

Submatrix $K$ is composed of $(1+n_T)$ submatrices $K^{A}$ for $A \in \{\partial H, \partial F_i\}$. Each of the $K^{A}$ has size $3\times 3N_A$, and its $3\times N_A$ components $K^{A}_{il,j}$ are defined as
\begin{equation}
K^{A}_{il,j} :=\int_A [(y-x_S) \wedge \vec{e}_i]_j \phi_l(x) \, \mathrm{d}x,
\end{equation}
for $i,j=1,2,3$ and $l=1,\dots,N_A$. In this case, the structure of matrix $K^A$ is block-antisymmetric.

Vector $I(\omega)$ is composed of $(n_T)$ non-zero subvectors $I(\omega)^A$ for $A \in \{\partial F_i\}$, and one zero subvector for $A= \partial H$. Each of the non-zero subvectors $I(\omega)^{A_i}$ is subdivided into $N_A$ subvectors of size $3$, named $I(\omega)^{A_i}_{j,l}$ defined as
\begin{equation}
I(\omega)^{A_i}_{j,l} = \int_{\partial {A_i}} [(x-x^{A_i}) \wedge \omega \vec{e}_1^{A_i}]_j\phi_l(x) \, \mathrm{d}x,
\end{equation}
for $j=1,2,3$ and $l=1,\dots,N_A$.

\end{document}